\documentclass[preprint,tightenlines,aps,floatfix,showpacs]{revtex4-1}  

\pdfoutput=1

\usepackage{amsmath}    
\usepackage{amssymb}
\usepackage{amsfonts}
\usepackage{graphicx}   
\usepackage{subfigure}  
\usepackage{hyperref}   
\usepackage{bm}
\usepackage{natbib}

\hypersetup{pdfauthor={C.~Bracher and A.~Gonzalez},
            pdftitle={Propagation of charged particle waves in a uniform magnetic field},
            pdfsubject={Submitted to Physical Review A (January 2012)}}

\begin{document}

\title{Propagation of charged particle waves in a uniform magnetic field}

\date{\today}

\author{Christian Bracher}
\email{bracher@bard.edu}
\affiliation{Physics Program, Bard College, PO Box 5000, Annandale-on-Hudson, NY 12504}

\author{Arnulfo Gonzalez}
\affiliation{Mathematics \& Science Department, Golden West College, Huntington Beach, CA 92647}

\begin{abstract}
This paper considers the probability density and current distributions generated by a point-like, isotropic source of monoenergetic charges embedded into a uniform magnetic field environment.  Electron sources of this kind have been realized in recent photodetachment microscopy experiments.  Unlike the total photocurrent cross section, which is largely understood, the spatial profiles of charge and current emitted by the source display an unexpected hierarchy of complex patterns, even though the distributions, apart from scaling, depend only on a single physical parameter.  We examine the electron dynamics both by solving the quantum problem, i.~e., finding the energy Green function, and from a semiclassical perspective based on the simple cyclotron orbits followed by the electron.  Simulations suggest that the semiclassical method, which involves here interference between an infinite set of paths, faithfully reproduces the features observed in the quantum solution, even in extreme circumstances, and lends itself to an interpretation of some (though not all) of the rich structure exhibited in this simple problem.
\end{abstract}


\maketitle

\section{Introduction}
\label{sec:Intro}

The classical cyclotron motion of a charge $q$ in a homogeneous magnetic field $\bm{\mathcal B}$ is a simple textbook problem.  Hence, it may come as a surprise that electron waves emerging from an isotropic point source into the magnetic field environment, which are analyzed in this article, should display any behavior of interest.  Indeed, only a few papers have been devoted to the subject in the literature.

Experimentally, such sources can be realized by the interaction of negatively charged ions with monochromatic laser light provided the photon energy closely matches the binding energy (affinity) of the excess electron (photodetachment threshold).  In this regime, the De Broglie wavelength of the emitted electron is large compared to the size of the emiting ion, and the absence of a long-range interaction between the photoelectron and the remaining neutral atomic core means that the dynamics of the emitted electron wave can be controlled externally, using applied electric and magnetic fields.  Experimentally, modulations of the overall photodetachment rate of negative ions in magnetic traps near threshold were first observed by Blumberg et al.\ and subsequently explained using perturbation theory \cite{Blumberg1978a,Blumberg1979a,Greene1987a}; recently, the technique has been used to observe the fine structure and Zeeman splittings in negative ions \cite{Yukich2011a}.  Similar experiments measuring the influence of an electric field on the photodetachment cross section \cite{Bryant1987a,Gibson1993a,Gibson1993b,Gibson2001a} were performed following theoretical investigations that predicted strong modulations of the detachment rate with photon energy \cite{Slonim1976a,Fabrikant1981a,Wong1988a}.  In semiclassical terms, the observed behavior can be ascribed to the interference of the outgoing electron wave with the electron returning to the emitting ion.  This intuitive trajectory-based picture is known as closed orbit theory \cite{Du1988a,Du2004a}.  The detachment cross section in combined electric and magnetic fields was addressed in numerous theoretical papers \cite{Du1989a,Fabrikant1991a,Peters1993a,Peters1994a,Peters1997a,Liu1997a}, and explored in at least one experiment involving parallel fields \cite{Yukich2003a}.

Rather than the total cross section, represented by the photocurrent, we are mainly interested in the spatial characteristics of the electron wave emitted by the point source, which include both its density distribution in space, and its experimentally more accessible current profile.  Due to the force they exert on the electron, external electric and magnetic fields profoundly alter the simple spherical wave pattern of a free electron source.  Guided by the classical electron trajectories, the waves refract and fold over under the influence of the fields.  Generally, the electron can travel along more than one classical orbit from the atomic source to a given destination in space; the waves associated with these multiple paths overlap and cause marked interference.  Although approximate, semiclassical methods \cite{Berry1972a,Maslov1981a,Delos1986a} based on these trajectories present a valuable tool in the interpretation of features observed in the exact quantum solution of the problem.  For a point source of electrons with energy $E$, located at the position $\mathbf r'$, their wave function is given by the energy Green function $G(\mathbf r, \mathbf r'; E)$, a particular solution of the stationary Schr\"odinger equation in the external potentials that represents an outgoing wave in the vicinity of the source \cite{Economou1983a,Kramer2002a}.   These Green functions can be found in analytical form for simple field configurations, including a homogeneous electric or magnetic field \cite{Dalidchik1976a,Bracher1998a,Bakhrakh1971a,Gountaroulis1972a,Dodonov1975a}.  In an electric field environment, charges classically follow free-fall parabolas, and their resulting spatial distribution features a regular interference pattern that is conveniently explained as interference between two classical paths, akin to the double slit model \cite{Fabrikant1981a,Demkov1982a,Du1989b,Golovinskii1997a}.  This theoretical prediction was confirmed experimentally by Blondel et al.\ \cite{Blondel1996a,Blondel1999a} who recorded the spatial current profile of electrons stripped from a negative ion beam using a tunable laser in an external field, thus imaging their wave function.  Because the interference pattern is exquisitely sensitive to the electron energy $E$, their technique, photodetachment microscopy, has become the gold standard in the precision measurement of electron affinities \cite{Blondel2001a,Blondel2005a,Pelaez2009a,Pelaez2011a}.  More recently, a similar analysis was performed for the case of parallel electric and magnetic fields.  The simultaneous action of both forces leads to more complex trajectory fields with an adjustable number of interfering paths, and correspondingly more involved interference patterns that are nevertheless in excellent agreement with the exact quantum solution \cite{Kramer2001a,Bracher2006a,Bracher2006b}.  First experimental images of the electron distribution observed in photodetachment in parallel fields have recently become available \cite{Blondel2010a}.

\begin{figure}[p]
\includegraphics[width=0.65\textwidth]{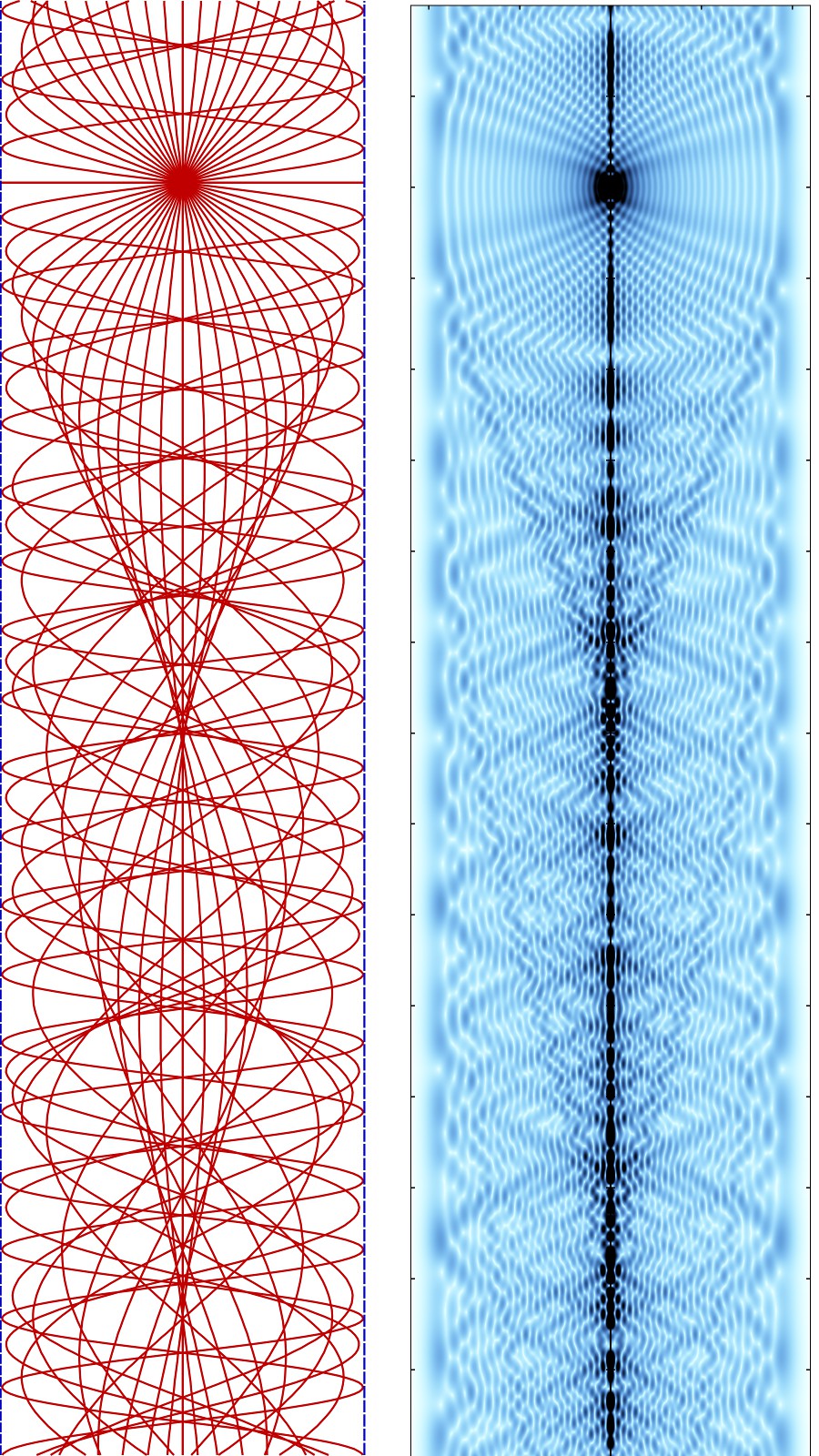}
\caption{\label{fig:Intro}
Electron wave emitted by an isotropic point source (near top of image) in a magnetic field.  Left panel:  Trajectory field, plotted as the radial distance $\hat\rho$ (horizontal) vs.\ the distance in field direction $\hat z$ (vertical) (dimensionless units).  Right panel:  Quantum calculation of the charge density along a radial section for an electron energy $E = 50 \hbar\omega_L$ (where $\omega_L$ denotes the Larmor frequency of the electron), corresponding to scattering into 24 open Landau channels.  Dark spots correspond to high density.}
\end{figure}

In contrast, the features of the electron wave propagating from a point source in the presence of a purely magnetic field have received little attention.  Naively, one might expect a less complex pattern due to the simplicity of the applied field, but it turns out that the electron waves now display a richer variety of structure than in the cases mentioned above.  To illustrate this, we display a sample plot of the electron density in Figure~\ref{fig:Intro} that shows a variety of interference phenomena at different scales.  These features have been partly predicted by Berry \cite{Berry1981a} who pointed out the presence of an infinite sequence of caustics, turning surfaces of classical motion that confine the classical trajectory fields \cite{Berry1972a,Nye1999a}, in this problem.  In the naive semiclassical model, the wave function diverges at these singular points \cite{Thom1975a,Berry1976a,Poston1978a,Berry1981b,Nye1999a}; this unphysical behavior can be corrected using uniform approximations \cite{Berry1976a,Schulman1981a,Nye1999a} adapted to the specific character of the singularity.  In his paper, Berry \cite{Berry1981a} predicted the amplification of the electron waves near the caustics, but stopped short of calculating the electronic wave function itself.  Our aim here is to systematically examine the properties of the electron wave from a classical, semiclassical, and quantum mechanical perspective, following the leads of a former study of the parallel field configuration \cite{Bracher2006b}.  We find that much of the added complexity in the magnetic case owes to the fact that the electron now can travel (if at all) along an infinite number of different classical paths from the source to any given destination, and that the electron drifts without acceleration along the magnetic field direction $\bm{\mathcal B}$.  As a result, the electronic wave combines properties of open scattering systems and closed problems with bound motion, best exemplified by quantum billiards \cite{Heller1984a,Berry1989a,Gutzwiller1990a,Jensen1992a}, although some of the features we observe, such as ``backflow'' of the electron toward the source, defy easy explanations.  Another observation of our study is the extreme resiliency of the semiclassical method which accurately renders the quantum solution even under adverse conditions.
%
%

We briefly comment on the structure of this paper.  In Section~\ref{sec:Classical}, we review the classical cyclotron motion of a charge $q$ in a magnetic field $\bm{\mathcal B}$ and show that the field of trajectories emitted by the source, apart from simple scalings in size and time, is universal.  In preparation of the semiclassical approach, we devise a method to identify the trajectories leading from the source to a given destination point; as a by-product, we obtain an alternative, simpler form of the caustic structure in this system first described by Berry \cite{Berry1981a}.  Studying the charge density, we find that the problem under consideration is classically ill-defined:  Interference and cancellation of waves is indispensible for the continuous operation of the source in a magnetic field. --- The quantum mechanical solution of our problem is the subject of Section~\ref{sec:Quantum}.  Apart from scaling, it depends on only a single dimensionless energy parameter $\epsilon$.  We determine the energy Green function by separation of the Hamiltonian in a bound, ``magnetic'' part perpendicular to $\bm{\mathcal B}$, and free propagation along the field direction, and obtain a generally rapidly converging series representation of $G(\mathbf r,\mathbf r'; E)$ that can be interpreted as a sum over discrete open and closed ``channels'' related to the quantized Landau levels in the magnetic field, but bears no formal resemblance to the classical description.  To bridge the divide in an attempt to understand the features of the quantum solution, we take up a semiclassical, trajectory-based approach in Section~\ref{sec:Semi}, and construct approximate solutions for the Green function and the associated charge and current densities $n(\mathbf r)$, $\mathbf j(\mathbf r)$ in the form of infinite series, where each term represents a possible trajectory of the electron.  In the semiclassical picture, the Landau levels arise as singularities in the spectrum caused by constructive interference of these paths, as pointed out already by Berry \cite{Berry1981a}.  To correct the failure of the naive semiclassical approximation near the caustics, we also deploy a uniform approximation based on Airy functions \cite{Abramowitz1965a,Olver2010a}.  In Section~\ref{sec:Results}, we study the charge and current density distributions for different values of the energy parameter $\epsilon$.  We find that the cyclotron motion confines the charge to a cylindrical tube aligned with the magnetic field, with the source at its center; the electron streams away from the source in a symmetrical pattern in both directions. As Figure~\ref{fig:Intro} illustrates, the charge and current density profiles display remarkably rich structure in all cases except for the lowest values of $\epsilon$:  An underlying finely detailed interference pattern is modulated in intensity along the caustics, as predicted by Berry \cite{Berry1981a}, but also contains embedded arc-like superstructure that superficially resembles the ``quantum scars'' observed in closed systems \cite{Heller1984a,Berry1989a,Wilkinson1996a}.  In the absence of an electric field, the density and current patterns are distinct, even though they often show common features.  However, near the energy thresholds where a new Landau channel opens (odd integer values of $\epsilon$), we encounter strongly diverging charge and current density profiles, coupled with a dramatic increase in the charge density associated with a simple interference pattern \footnote{Experimentally, finite coherence time and the resulting limited spectral resolution will ``soften'' these divergences and alter the near-threshold interference patterns in interesting ways.  These effects could be examined by propagating wave packets in a magnetic field, but such a study is outside the scope of this paper.}.  Furthermore, ``backflow,'' the presence of extended regions in space where the electron is moving toward the source instead of flowing away, as one would naively expect, is prevalent in this regime. --- In the concluding remarks, we summarize our results, and briefly explore the feasibility of experimental confirmation.

\section{Classical Dynamics}
\label{sec:Classical}

We first describe the motion of a charge streaming from the source (which we will identify with the coordinate origin $\mathbf r' = \mathbf 0$) from a classical viewpoint, and show how to obtain the trajectories leading to any given destination $\mathbf r$.  For simplicity, we will align the magnetic field vector $\bm{\mathcal B} = \mathcal B \hat e_z$ with the $z$--axis, and keep a generic value for the particle charge $q$ and mass $m$ in our calculation.  We assume that $q>0$, but point out that none of our results, with the exception of the azimuthal angle $\phi(t)$ in Eqs.~(\ref{eq:Classical7}) and (\ref{eq:Classical8}) \footnote{A negative charge rotates counterclockwise in the $x$-$y$ plane.}, will change for a negative charge if $q$ is replaced by its absolute value $|q|$.  (For details concerning the classical dynamics, see one of the authors' (A.~G.) Masters thesis \cite{Gonzalez2009a}.  A formally similar analysis applies to the problem of electron motion in parallel fields, laid out in Ref.~\cite{Bracher2006a,Bracher2006b}, where the purely magnetic case is obtained in the limit $\eta = v_0\mathcal B/\mathcal E \rightarrow \infty$.)

\subsection{Cyclotron motion}

We first find the trajectories $\mathbf r(t)$ for a particle emitted under a spherical angle $(\theta_0,\phi_0)$ from the source.  We tackle this problem using the principle of least action,
\begin{equation}
\label{eq:Classical1}
S(\mathbf r, t; \mathbf 0, 0) \,=\, \min_{\mathbf r(t)} \int_0^t dt' \left(\frac m2 \dot{\mathbf r}^2 + q\dot{\mathbf r} \cdot \mathbf A(\mathbf r) \right)\;,
\end{equation}
using the minimal coupling Lagrangian $\mathcal L\left(\mathbf r(t), \dot{\mathbf r}(t)\right)$ with a suitable vector potential $\mathbf A(\mathbf r)$.  We choose a gauge for $\mathbf A(\mathbf r)$ that conforms to the inherent cylindrical symmetry:
\begin{equation}
\label{eq:Classical2}
\mathbf A(\mathbf r) \,=\, \frac12 (\bm{\mathcal B} \times \mathbf r) \,=\, \frac{B}2 \left( -y, x, 0 \right)^T \;,
\end{equation}
and express $\mathcal L\left(\mathbf r(t), \dot{\mathbf r}(t)\right)$ in cylindrical coordinates $(\rho, z, \phi)$:
\begin{equation}
\label{eq:Classical3}
\mathcal L\left(\mathbf r, \dot{\mathbf r}\right) \,=\, \frac m2 \left(\dot\rho^2 + \dot z^2 + \rho^2\dot\phi^2 \right) + \frac{qB}2 \rho^2 \dot\phi \;.
\end{equation}
Since the coordinates $z$ and $\phi$ are cyclic, and the Hamiltonian is conservative, there are three conserved quantities of motion:  $p_z = m\dot z = \text{const.}$ indicates free particle motion along the field axis, while the momentum conjugate to $\phi$ is the constant canonical angular momentum $L_z$ of the charge:
\begin{equation}
\label{eq:Classical4}
L_z \,=\, \partial \mathcal L/\partial \dot\phi \,=\, m\rho^2\dot\phi + \frac{qB}2 \rho^2 \,=\, \text{const.}
\end{equation}
For a trajectory traversing the origin $\rho=0$, $L_z$ must be zero, so the trajectory undergoes uniform rotation in the $x$-$y$ plane with an angular velocity $\omega_L$:
\begin{equation}
\label{eq:Classical5}
\left| \dot\phi \right| \,=\, \omega_L \,=\, \frac{q\mathcal B}{2m} \;,
\end{equation}
the Larmor frequency of the problem \footnote{The familiar cyclotron frequency $\omega_C = 2\omega_L$ describes the rotating charge as seen from the center of its orbit.  Here, we study the angular motion with regard to the source, which is a point on the orbit itself.}.  Finally, the conserved energy $E$, given by the Hamilton function of the system:
\begin{equation}
\label{eq:Classical6}
\mathcal H(\mathbf r, \mathbf p) \,=\, \dot{\mathbf r} \cdot \mathbf p - \mathcal L \,=\, \frac m2 \left(\dot\rho^2 + \dot z^2 + \rho^2\dot\phi^2 \right) \,=\, \text{const.},
\end{equation}
is simply the kinetic energy $E = \frac12 m {\dot{\mathbf r}}^2$ of the charge, which therefore travels at a constant speed $v_0$.  (In cartesian coordinates, the Hamilton function takes the form $\mathcal H = \frac1{2m} (\mathbf p - q\mathbf A)^2$, which is useful for the quantum description; see Appendix.)

Solving the equations of motion for a trajectory emitted under polar angle $(\theta_0,\phi_0)$ yields the familiar helical orbit:
\begin{equation}
\label{eq:Classical7}
\rho(t) = \frac {v_0}{\omega_L} \sin{\theta_0} |\sin(\omega_L t)| \;, \quad
z(t) = v_0 t \cos{\theta_0} \;, \quad
\phi(t) = \phi_0 - \omega_L t \;.
\end{equation}
Note that all trajectories share the same rotational motion.  It is therefore sufficient to study the charge dynamics in the $\rho$-$z$ space, where the path becomes a simple sine curve whose frequency and amplitude depend on the polar emission angle $\theta_0$. (It is convenient to dispense of the absolute value in Eq.~(\ref{eq:Classical7}), and formally allow negative values for $\rho$.)

We point out that the trajectory field (\ref{eq:Classical7}) is universal in the sense that for any given set of parameters, the motion of the charge differs only in its range and speed.  We measure distances $\hat\rho = \omega_L \rho / v_0$, $\hat z = \omega_L z / v_0$ in units of the maximum cyclotron orbit diameter $v_0 / \omega_L$, and introduce a dimensionless time $\tau = \omega_L t$.  The charge then completes a cyclotron orbit of radius $\sin\theta_0$ within a time period $\tau_\text{cyclo} = \pi$:
\begin{equation}
\label{eq:Classical8}
\hat\rho(\tau) = \sin{\theta_0} \sin\tau \;, \quad
\hat z(\tau) = \tau \cos{\theta_0}  \;, \quad
\phi(\tau) = \phi_0 - \tau \;.
\end{equation}

\subsection{Finding trajectories}

While Eq.~(\ref{eq:Classical8}) details all trajectories, we are interested in finding those paths that lead to a given target $\mathbf r$ with dimensionless coordinates $(\hat\rho, \hat z,\phi)$.  Eliminating the emission angle $\theta_0$ in the set of equations (\ref{eq:Classical8}) yields implicit solutions in terms of a function $\epsilon(\tau)$ of the time of flight $\tau$:
\begin{equation}
\label{eq:Classical9}
\hat\epsilon(\tau) \,=\, \frac{\hat\rho^2}{\sin^2\tau} + \frac{\hat z^2}{\tau^2} \,=\, 1 \;.
\end{equation}
It is easily verified that there is exactly one trajectory $\mathbf r(t)$ that connects the source with the destination $(\rho, z)$ in a given time $T$ if one allows the energy of the charge to vary; $\hat\epsilon(\tau)$ (\ref{eq:Classical9}) then indicates the energy of this trajectory in units of the fixed source energy $E = \frac12 mv_0^2$.  We therefore denote $\hat\epsilon(\tau)$ the {\sl energy function} in this problem, and the condition (\ref{eq:Classical9}) is now seen to pick the trajectories of the ``correct'' energy.
%
%
\begin{figure}[t]
\includegraphics[width=0.75\textwidth]{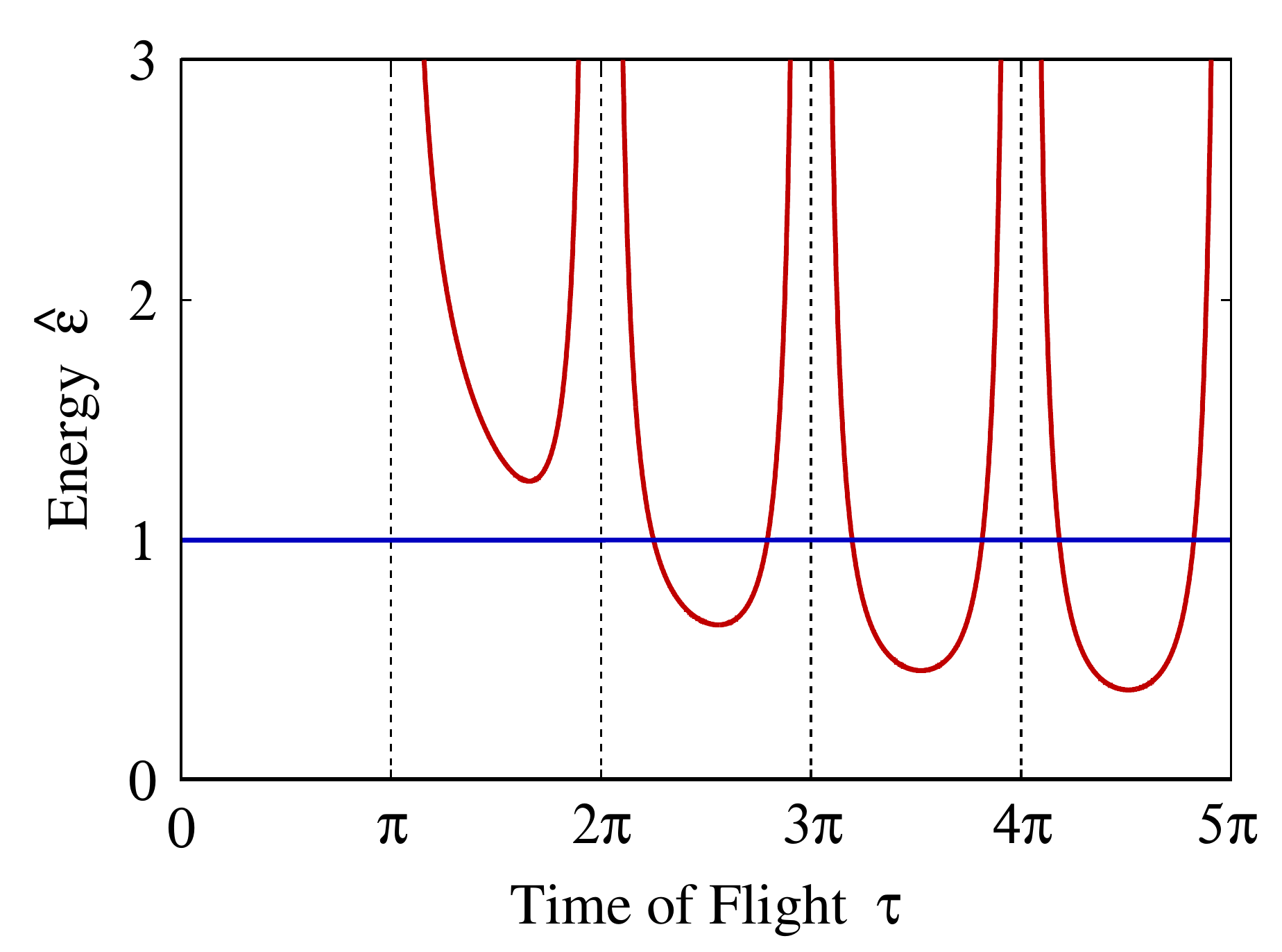}
\caption{\label{fig:Epsilon}
Graph of the dimensionless energy function $\hat\epsilon(\tau)$ (red (gray) curve).  The intersections with the horizontal line $\hat\epsilon = 1$ (blue) correspond to the possible times of flight $\tau$.  Starting from some initial value $\nu_0$ (here, $\nu_0=2$), we find two solutions per cyclotron interval $\nu\pi < \tau < (\nu+1)\pi$, implying an infinite number of trajectories.  (Parameters used: $\rho = 0.5$, $\hat z = 5$.)}
\end{figure}

Figure~\ref{fig:Epsilon} shows a plot of $\hat\epsilon(\tau)$ that reveals some important features of this function:  It is strictly positive and diverges at every integer multiple $\tau = \nu\pi$ of the cyclotron period $\tau_\text{cyclo}$, giving the graph the appearance of a succession of disjoint U-shaped curves in each interspersed region.  The minima of these curves, located at the roots of $\hat\epsilon'(\tau)$ \footnote{There is only one such minimum in each interval because $\hat\epsilon(\tau)$ is a convex function:  $d^2{\hat\epsilon}/d\tau^2 = 2\rho^2(1 + 2\cos^2\tau)/\sin^4\tau + 6\hat z^2/\tau^4$ is strictly positive.}:
\begin{equation}
\label{eq:Classical11}
\frac{d\hat\epsilon}{d\tau}\,(\tau) \,=\, -2 \left( \frac{\hat\rho^2 \cos\tau}{\sin^3\tau} + \frac{\hat z^2}{\tau^3} \right)\;,
\end{equation}
form a monotonically decreasing sequence that tends toward ${\hat\rho}^2$ as the number $\nu$ of the cyclotron interval tends toward infinity.  Hence, classical trajectories do not reach beyond the cylindrical region $\hat\rho \geq 1$. (We note, however, that the analytic continuation of $\hat\epsilon(\tau)$ formally admits complex-valued solutions for the time of flight $\tau$.  These lead to complex ``ghost orbits'' that lack physical reality from the classical viewpoint, but acquire meaning in the semiclassical description as exponentially suppressed ``tunneling trajectories'' that make the classically forbidden region accessible to waves.  We will use them in our simulations in this paper.)

Conversely, inside the cylinder $\hat\rho < 1$, there will always be an {\sl infinite} set of trajectories solving our problem.  Starting from some initial cyclotron orbit $\nu_0$, the graph of $\hat\epsilon(\tau)$ will intersect $\hat\epsilon = 1$ twice in every subsequent interval $\nu\geq \nu_0$, corresponding to a pair of ``fast'' and ``slow'' trajectories that undergo $\nu$ complete cyclotron orbits before arriving at $(\hat\rho, \hat z)$.  Physically, these trajectories form ever more tightly wound helices leading from source to destination.

In general, the transcendental equation (\ref{eq:Classical9}) has no analytic solutions, so we rely on numerical methods to find the times of flight $\tau_\nu$ and corresponding trajectories $\mathbf r_\nu(\tau)$ of the charge.  Our fast and reliable scheme is based on Newton's method.  Since all other physical quantities of interest can be expressed as functions of the time of flight $\tau$, finding the roots in Eq.~(\ref{eq:Classical9}) is the only instance where we resort to numerical computation.  We note, however, that in the limit $\nu \rightarrow \infty$, the times $\tau$ asymptotically approach a regular sequence:
\begin{equation}
\label{eq:Classical12}
\tau \,\sim\, \nu\pi + \arcsin{\hat\rho} \quad \text{and} \quad
\tau \,\sim\, (\nu+1)\pi - \arcsin{\hat\rho} \qquad (\nu \rightarrow\infty) \;.
\end{equation}
This approximation is helpful when studying the convergence properties of the source problem from the classical and semiclassical perspective.

\subsection{Caustics}

Although an infinitude of classical paths will connect any destination point $\mathbf r$ (with $\hat\rho < 1$) to the origin, the qualitative character of the solutions generally differs among these points, because they do not all share the same minimum amount $\nu_0$ of cyclotron orbits required to reach the destination.  For each value of $\nu_0$, the set of destinations forms a separate manifold, and a pair of trajectories is gained or lost whenever $\mathbf r$ crosses over its boundary to the ``neighboring'' manifolds requiring $\nu_0 \pm 1$ cyclotron orbits.  These singularities in the trajectory field, where the smooth mapping between trajectories $\mathbf r(t)$ and the target $\mathbf r$ breaks down, are known as the caustic set \cite{Nye1999a,Berry1981b}.

\begin{figure}[t]
\includegraphics{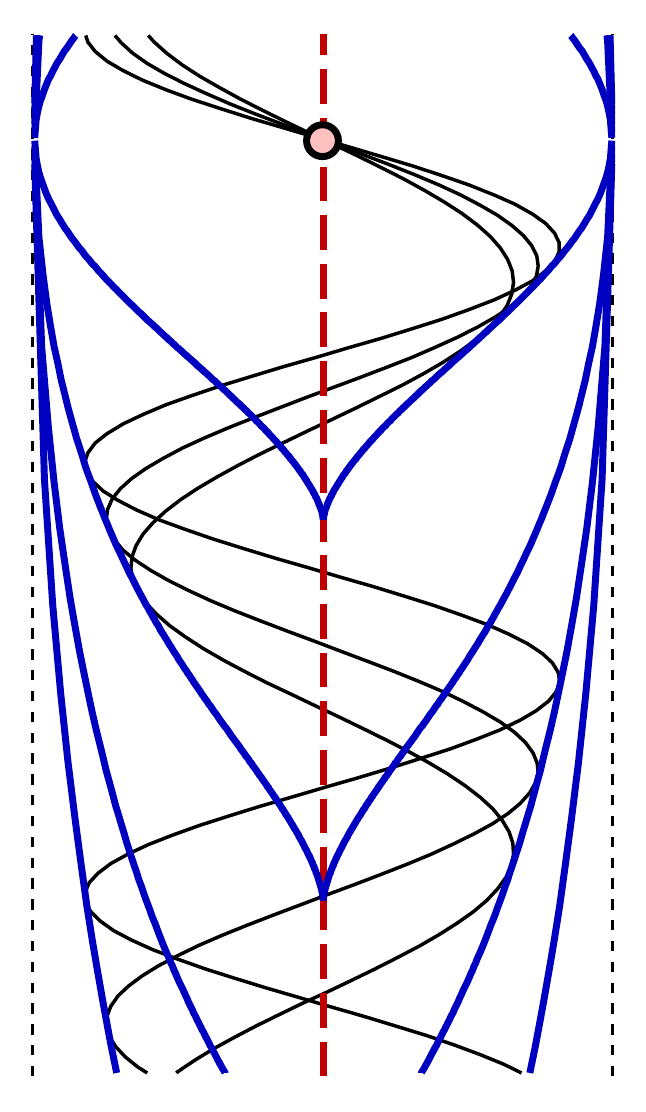}
\caption{\label{fig:Caustics}
Caustic pattern in the $\hat\rho$-$\hat z$ plane for the motion of a charge in a magnetic field.  During their $\nu$th cyclotron orbit, the trajectories (thin curves) trace out the $\nu$th caustic surface (bold blue curves).  They form an onion-like nested set.  Besides these surfaces, a singularity in the trajectory field occurs along the symmetry axis $\hat\rho = 0$ (dashed red line).}
\end{figure}

In our problem, two distinct types of caustic points are present \cite{Berry1981a}: There is an infinite sequence of rotationally symmetric, onion-shaped caustic surfaces that are nested inside each other.  These surfaces are related to the cyclotron motion of the charge, and the $\nu$th surface \footnote{We enumerate the caustic surfaces starting with $\nu=0$.} represents the set of destinations where trajectories requiring $\nu$ prior complete loops go in and out of existence. For a quantitative study, we employ the scheme for finding solutions (Figure~\ref{fig:Epsilon}) introduced above.  Accordingly, the $\nu$th caustic is comprised of those points where the intersection of the energy functional $\hat\epsilon(\tau)$ with the line $\hat\epsilon = 1$ in the $\nu$th cyclotron interval is lifted, so $\hat\epsilon = 1$ is tangential to the graph of $\hat\epsilon(\tau)$.  Thus, the caustic is implicitly defined through the pair of equations:
\begin{equation}
\label{eq:Classical13}
\hat\epsilon(\tau) \,=\, 1 \quad \text{and} \quad \frac{d{\hat\epsilon}}{d\tau}\,(\tau) \,=\, 0 \;,
\end{equation}
where $\nu\pi < \tau < (\nu+1)\pi$.  Due to the transcendental nature of the equations (\ref{eq:Classical9}) and (\ref{eq:Classical11}) the shape of the caustic cannot be represented by a closed expression $\hat \rho(\hat z)$.  However, it is possible to reorder them, and obtain a parametrization of the caustics by the time of flight $\tau$:
\begin{equation}
\label{eq:Classical14}
\hat\rho(\tau) \,=\, \sqrt{\frac{\sin^3\tau}{\sin\tau - \tau\cos\tau}} \;, \qquad
\hat z(\tau) \,=\, \pm\sqrt{\frac{\tau^3\cos\tau}{\tau\cos\tau - \sin\tau}} \;,
\end{equation}
where $\tau$ runs through the interval $(\nu + \frac12)\pi \leq \tau \leq (\nu+1)\pi$.  Note that at the lower limit $\tau = (\nu+\frac12)\pi$ of this interval, one finds $\hat\rho=1$ and $\hat z=0$.  As $\tau$ increases, $\hat\rho$ shrinks and $|\hat z|$ increases, until the caustic ends in a cusp-like structure at the symmetry axis ($\hat\rho =0$) at the locations $\hat z = \pm (\nu+1)\pi$, the distance traveled in $\nu+1$ complete orbits.  Therefore, the caustics are symmetric surfaces stacked inside each other, and are joined at the circle $\hat\rho = 1$ in the $x$-$y$ plane.  (An alternative, more complicated parametrization using the emission angle $\theta_0$ has been proposed by Berry \cite{Berry1981a}.)  The pattern is shown in Figure~\ref{fig:Caustics}.  To illustrate the relation between caustics and orbits further, we include a few trajectories in the graph (thin line).  They cross the symmetry axis $\nu$ times, corresponding to $\nu$ complete cyclotron orbits, before being ``reflected'' off the $\nu$th caustic surface.  Note that like the trajectory field, the caustic set has a universal shape, only subject to scaling.
%
%

Besides these ``fold'' caustics \cite{Nye1999a,Berry1976a,Poston1978a,Berry1981b}, there is an unrelated second type of caustic that extends along the entire symmetry axis ($\hat\rho = 0$) of the problem.  Along the symmetry axis, all trajectories emitted under a fixed polar angle $\theta_0$ will converge again in a single point, irrespective of their azimuthal emission angle $\phi_0$.  Again, this symmetry lifts the smooth mapping between trajectories and destination point, and leads to the formation of the caustic. --- Finally, we emphasize that the singularities are associated with the trajectory {\sl field} only; the propagation of a charge along a given trajectory $\mathbf r(t)$ is always smooth, even when it traverses a caustic point.

\subsection{Classical density}

Having identified the individual trajectories $\mathbf r_\nu(t)$ that lead to the destination $\mathbf r$, we need to find the relative weight carried by each of these paths in order to fully establish the classical and semiclassical dynamics of the charge in a uniform magnetic field.  To this end, we examine how a bundle of ``neighboring'' trajectories akin to $\mathbf r_\nu(t)$ spreads as it travels from the source to the target.

Consider an infinitesimally small volume $d{\hat V} = \rho d\rho dz d\phi = (v_0 / \omega_L)^3 \hat\rho d\hat\rho\,d\hat z\,d\phi$ surrounding the destination $\mathbf r$.  Assuming that the trajectory $\mathbf r_\nu(t)$ arriving at $\mathbf r$ is emitted under angles $(\theta_0,\phi_0)$ and has a time of flight $\tau$ to the destination, the trajectories leading into the surrounding volume will have a spread $(d\theta_0,d\phi_0)$ in emission angle, and $dt = d\tau / \omega_L$ in the time of flight.  Assuming isotropic characteristics of the source, and a total flux of $J$ charges per second, the number of particles emitted within this spherical angle and time interval is:
\begin{equation}
\label{eq:Classical15}
dN_\nu \,=\, \frac J{4\pi\omega_L} \, \sin\theta_0 d\theta_0 \, d\phi_0 \, d\tau \;.
\end{equation}
Since the number of particles is conserved during propagation, and the source is stationary, $dN_\nu$ is also the number of charges populating the volume $dV$ that stem from the trajectory family associated with $\mathbf r_\nu(t)$. Hence, the density $n_\nu(\mathbf r) = dN_\nu/dV$ due to these trajectories is:
\begin{equation}
\label{eq:Classical16}
n_\nu(\mathbf r) \,=\, \frac{J \omega_L^2}{4\pi v_0^3} \, \frac{\sin\theta_0}{\hat\rho} \, \frac{\partial(\theta_0,\phi_0,\tau)}{\partial(\hat\rho,\hat z,\phi)} \,=\, \frac{J \omega_L^2}{4\pi v_0^3} \, \frac{\sin\theta_0}{\hat\rho} 
\left| \det \mathcal J \right|^{-1} \;,
\end{equation}
where $\mathcal J = \partial(\hat\rho,\hat z,\phi) / \partial(\theta_0,\phi_0,\tau)$ is the Jacobian matrix for the transformation (\ref{eq:Classical8}) relating the initial conditions (emission angle and time of flight) to the final position.  Its determinant is $\det \mathcal J = \sin^2\theta_0 \tau\cos\tau + \cos^2\theta_0 \sin\tau$.  Using Eq.~(\ref{eq:Classical8}), we substitute the emission angles via $\sin\theta_0 = \hat\rho/\sin\tau$ and $\cos\theta_0 = \hat z/\tau$:
\begin{equation}
\label{eq:Classical17}
n_\nu(\mathbf r) \,=\, \frac{J \omega_L^2}{4\pi v_0^3} \, \frac 1{\tau\sin^2\tau} \, \left| \frac{\hat\rho^2\cos\tau}{\sin^3\tau} + \frac{\hat z^2}{\tau^3} \right|^{-1} 
\,=\, \frac{J \omega_L^2}{2\pi v_0^3 \tau\sin^2\tau \left| d{\hat\epsilon}/ d\tau \right|} \;,
\end{equation}
by comparison with the energy functional in Eq.~(\ref{eq:Classical11}).  As an immediate consequence, we infer that the classical particle density $n_\nu(\mathbf r)$ diverges at the location of the source (where $\tau\rightarrow 0$), and at both types of caustics:  For $\sin\tau = 0$, i.~e.\ $\tau = \nu\pi$, the charge traverses the symmetry axis ($\hat\rho = 0$), while $d{\hat\epsilon}/d\tau = 0$ holds along the off-axis caustic surfaces according to the condition (\ref{eq:Classical13}).

Outside the caustic set, the density $n_\nu(\mathbf r)$ is finite for each individual trajectory $\mathbf r_\nu(t)$.  However, the {\sl total} classical particle density $n_\text{cl}(\mathbf r)$, obtained by summation over all paths leading to the destination, still diverges, as there is an infinite number of slow paths. The density contribution due to such a path undergoing a large number $\nu \rightarrow \infty$ of cyclotron orbits is approximately:
\begin{equation}
\label{eq:Classical18}
n_\nu(\mathbf r) \,\sim\, \frac {J \omega_L^2}{4\pi^2 v_0^3} \, \frac 1{\hat\rho\sqrt{1-\hat\rho^2}} \, \frac 1\nu  \qquad (\nu \rightarrow\infty)\;.
\end{equation}
(According to Eq.~(\ref{eq:Classical12}), $\tau \sim \nu\pi$, and $\sin\tau \approx \rho$.)  Summation over all $\nu$ therefore yields a divergent harmonic series, indicating an infinite classical density throughout the cylinder $\hat\rho < 1$.  This result indicates that the concept of a perfectly stationary source in a magnetic field is invalid from a classical perspective.  Remarkably, the quantum and semiclassical solutions to the problem do not suffer from this defect, as we shall demonstrate now.

\section{Solving the Quantum Problem}
\label{sec:Quantum}

We now turn our attention to the quantum mechanical description of the problem, and study the propagation of a wave under the time evolution operator $\mathcal U = \exp(-i\mathcal H t/\hbar)$, where $\mathcal H(\mathbf r, \mathbf p) = \frac 1{2m}[\mathbf p - q\mathbf A(\mathbf r)]^2$ is the Hamilton operator for a charge in a purely magnetic field.  Being mainly concerned with the results, we leave a sketch of the solution to the Appendix.

\subsection{Quantum mechanics with sources}

One conceptual difficulty that arises in the quantum problem is that the classical trajectory field (\ref{eq:Classical8}) describes particles spreading from the origin.  Thus, ``new'' particles are continuously generated there at a constant rate.  This is at odds with the conventional picture of quantum mechanics, where the equation of continuity $\partial_t n + \bm\nabla \cdot \mathbf j = 0$ for probability density $n(\mathbf r,t)$ and current density $\mathbf j(\mathbf r, t)$ implies that the number of particles is conserved.  One workaround to the problem modifies the stationary Schr\"odinger equation itself, and introduces an inhomogeneous source term into the equation.  For a point-like source of ``unit strength,'' the Schr\"odinger equation takes the form (\ref{eq:Appendix1}):
\begin{equation}
\label{eq:Quantum-1}
\left[E - \mathcal H(\mathbf r,\mathbf p)\right] \psi(\mathbf r) \,=\, \delta(\mathbf r) \;.
\end{equation}
A review of the source formalism, including spatially extended sources, is contained in Ref.~\onlinecite{Kramer2002a}.

Mathematically, Eq.~(\ref{eq:Quantum-1}) represents the concept of a Green function, and so we call the set of solutions $\psi(\mathbf r) = G(\mathbf r, \mathbf 0 ;E)$ the energy Green function of the charge in the magnetic field.  (See the Appendix for a discussion.)  We pick a wave function with a current field $\mathbf j(\mathbf r)$ that radially spreads from the source, matching the isotropic emission pattern we assumed for the classical trajectory field in Eq.~(\ref{eq:Classical15}).  (The source $\delta(\mathbf r)$ has no preferred direction in space, and therefore invariably has $s$--wave characteristics.  The point source formalism can be modified to comprise emission into $p$--waves and higher angular momenta \cite{Bracher2003a}, an extension we do not address in this paper.)  In a more traditional interpretation, this ``retarded'' solution represents the outgoing wave part of a scattering wave function solving the conventional Schr\"odinger equation.  Interestingly, the total particle current $J(E) = -\frac 2\hbar \Im[G(\mathbf0, \mathbf 0;E)]$ emitted by the point source can be directly read off the imaginary part of $G(\mathbf r,\mathbf 0;E)$ at the location of the source, as shown in the Appendix.

If the motion of the charge is restricted to a two-dimensional plane, the problem of propagation in a magnetic field admits a closed-form quantum solution \cite{Dodonov1975a,Bellandi1976a}, while in the three-dimensional problem, the Green function takes the form of an infinite series \cite{Gountaroulis1972a,Kramer2005a}, as outlined in the Appendix.

\subsection{The magnetic Green function}

In order to establish the energy Green function $G(\mathbf r,\mathbf 0;E)$ for a charge in the magnetic field in analytic form, we note that the Hamiltonian operator $\mathcal H(\mathbf r,\mathbf p) = \frac 1{2m} [\mathbf p - q\mathbf A(\mathbf r)]^2$ is a sum of two commuting parts, a perpendicular operator $\mathcal H_\perp$ which details the cyclotron motion in the magnetic field, and a longitudinal operator $\mathcal H_\|$ covering the free motion in field direction.  (This is analogous to the classical case, where the energies in transversal motion $E_\perp = \frac 12 mv_0^2\sin^2\theta_0$ and motion parallel to the field $E_\| = \frac 12 mv_0^2 \cos^2\theta_0$ are separately conserved.)  $G(\mathbf r,\mathbf 0;E)$ then can be expressed as a sum over products of the various ``transversal'' eigenfunctions of a charge in the magnetic field, which are arranged in Landau levels $E_l = (2l+1) \hbar\omega_L$ \cite{Moore1977a}, with the simple free-particle Green function in one dimension, evaluated for the matching energy $E_\| = E - E_l$.  A derivation of the magnetic Green function using this product approach is sketched in the Appendix.  (A similar procedure leads to the Green function for a charge in parallel electric and magnetic fields \cite{Fabrikant1991a,Kramer2001a,Bracher2006b}, from which it emerges in the (non-trivial) limit $\mathcal E \rightarrow 0$.)  For an overview of methods to calculate energy Green functions, we refer to Ref.~\cite{Kramer2005a}, which contains an alternative derivation of $G(\mathbf r,\mathbf 0;E)$ in the magnetic field environment.  Yet another approach is due to Gountaroulis \cite{Gountaroulis1972a}.

Unlike the classical trajectory field, the wave function $G(\mathbf r, \mathbf 0; E)$ is no longer universal, but depends on a dimensionless energy parameter $\epsilon$, measured in terms of the ground state energy $\hbar\omega_L$ of the charge in the magnetic field:
\begin{equation}
\label{eq:Quantum1}
\epsilon \,=\, E/\hbar\omega_L \;.
\end{equation}
Alternatively, the parameter $\epsilon$ can be interpreted as the ratio $\epsilon = 2\pi\rho_\text{cyclo} / \lambda$ of the classical cyclotron radius $\rho_\text{cyclo} = v_0/(2\omega_L)$ and the De Broglie wavelength of the charge $\lambda = 2\pi\hbar/(mv_0)$; large values of $\epsilon$ imply that the quantum length scale $\lambda$ is small compared to the classical length scale $\rho_\text{cyclo}$.  In terms of $\epsilon$ and the scaled coordinates $\hat\rho$ and $\hat z$, the Green function has the series representation (see Eq.~(\ref{eq:Appendix19}) in the Appendix):
\begin{equation}
\label{eq:Quantum2}
G(\mathbf r, \mathbf 0; E) \,=\, \frac{mk}{2\pi\hbar^2} e^{-\epsilon \hat\rho^2} \left( 
\sum_{2l+1 < \epsilon} L_l(2\epsilon \hat\rho^2) \frac{e^{2i\sqrt{\epsilon(\epsilon-2l-1)} |\hat z|}}{i\sqrt{\epsilon(\epsilon - 2l - 1)}} -
\sum_{2l+1 > \epsilon} L_l(2\epsilon \hat\rho^2) \frac{e^{-2\sqrt{\epsilon(2l+1 -\epsilon)} |\hat z|}}{\sqrt{\epsilon(2l + 1 - \epsilon)}} \right) \;,
\end{equation}
where $k=\sqrt{2mE}/\hbar$ is the wave number, $L_l(u)$ denotes a Laguerre polynomial, and the sum runs over all Landau levels ($l = 0,1,2,3,\ldots$).  As the individual terms generally drop exponentially once $l > \epsilon/2$, the series converges rapidly, and $G(\mathbf r, \mathbf 0; E)$ can be accurately evaluated numerically.  The only exception is the perpendicular plane $\hat z=0$ containing the source.

Finally, we point out that the Green function $G(\mathbf r, \mathbf 0; E)$ is not defined at the Landau levels $\epsilon_l = 2l+1$ itself.  In the vicinity of these energies, $G(\mathbf r, \mathbf 0; E)$ grows indefinitely.  Mathematically, this relates to the observation that the Green function as the configuration space representation of the resolvent operator $(E - \mathcal H)^{-1}$ will diverge at eigenenergies $E_l$ in the discrete spectrum of $\mathcal H$.  (See also the discussion in the Appendix.)  Although there are no bound states in the full scattering problem, the periodic motion of the charge in the perpendicular $x$-$y$ plane caused by the magnetic field, and the ensuing quantization of the energy $E_\perp$ in transversal motion into Landau levels, becomes manifest in the singularities of $G(\mathbf r, \mathbf 0; E)$.

\subsection{Quantum charge densities and currents}

From the energy Green function $G(\mathbf r, \mathbf 0; E)$ we obtain both the probability density of the electron in the emitted wave:
\begin{equation}
\label{eq:Quantum3}
n_\text{qm}(\mathbf r) \,=\, \left|G(\mathbf r, \mathbf 0; E)\right|^2 \;,
\end{equation}
and the probability density current $\mathbf j_\text{qm}(\mathbf r)$, which yields the rate $\mathbf j \cdot d\mathbf a$ at which charges would impinge on a detector area element $d\mathbf a(\mathbf r)$ (where $\mathbf a$ is the normal vector to the detector surface):
\begin{equation}
\label{eq:Quantum4}
\mathbf j_\text{qm}(\mathbf r) \,=\, \frac 1{2m} \left[G(\mathbf r, \mathbf 0; E)^* \left(\mathbf p - q\mathbf A(\mathbf r)\right) G(\mathbf r, \mathbf 0; E) + \text{c.c.}\right] \;.
\end{equation}
Unlike in the presence of an electric field $\bm{\mathcal E}$ \cite{Kramer2001a,Bracher2006b}, where continuous acceleration tends to align the current profile with the density profile far from the source, the charge and current distributions retain their individual character in a purely magnetic environment, so the vector field $\mathbf j_\text{qm}(\mathbf r)$ warrants a detailed study.

Given the symmetry of the problem, it is convenient to extract the radial component $j_\rho$ and a component $j_z$ parallel to the field:
\begin{equation}
\label{eq:Quantum5}
j_z(\mathbf r) \,=\, \frac \hbar m \Im\left[G(\mathbf r, \mathbf 0; E)^* \partial_z G(\mathbf r, \mathbf 0; E)\right]
\;, \qquad j_\rho(\mathbf r) = \frac \hbar m \Im\left[G(\mathbf r, \mathbf 0; E)^* \partial_\rho G(\mathbf r, \mathbf 0; E)\right] \;.
\end{equation}
Since $G(\mathbf r, \mathbf 0; E)$ is a function of $\hat\rho$ and $\hat z$ only, the remaining azimuthal component $j_\phi$ is related to the vector potential term in Eq.~(\ref{eq:Quantum4}), and therefore proportional to the particle density $n_\text{qm}(\mathbf r)$:
\begin{equation}
\label{eq:Quantum6}
j_\phi(\mathbf r) \,=\, - v_0 \hat\rho \left|G(\mathbf r, \mathbf 0; E)\right|^2 \;,
\end{equation}
indicating uniform rotation of the particle wave with the Larmor frequency $\omega_L$, familiar from the classical dynamics (\ref{eq:Classical7}).  We will examine these quantities further in Section~\ref{sec:Results}.

Finally, the total current $J(E)$ can be read off the Green function.  Applying relation (\ref{eq:Appendix3}) in the Appendix, we find that only the first sum in Eq.~(\ref{eq:Quantum2}), corresponding to the open scattering channels, contributes to the current:
\begin{equation}
\label{eq:Quantum7}
J(E) \,=\, J_\text{free}(E) \sum_{0 < 2l+1 < \epsilon}  \frac 1{\sqrt{\epsilon(\epsilon - 2l - 1)}} \;.
\end{equation}
Here, $J_\text{free}(E) = mk/(\pi\hbar^3)$ is the current (\ref{eq:Appendix8}) emitted by a free particle source in the absence of the magnetic field.  Note that the total current $J(E)$, like $G(\mathbf r,\mathbf 0;E)$ itself, diverges at the energies $\epsilon_l = 2l+1$, i.~e., whenever a new scattering channel ``opens'' \footnote{Note that the singularity in Eq.~(\ref{eq:Quantum7}) is integrable, and thus will not affect experimental measurements of the total current with their necessarily finite energy resolution.}.  As our sample calculations will show, the simplicity of the result (\ref{eq:Quantum7}) belies the intricate and unusual structure present in the charge and current distributions. 

\section{Semiclassical Analysis}
\label{sec:Semi}

In order to gain insight into the features of the quantum density and current distributions, we now embark on a semiclassical study of electron dynamics in the magnetic field, based on the helical trajectories we identified in Section~\ref{sec:Classical}.  While not an exact method, we find that the semiclassical model faithfully reproduces the quantum results, in particular when used in conjunction with the uniform approximation which lifts the singularities of the semiclassical method that occur at the caustics. 

\subsection{The semiclassical wave function}

To start, we assemble the semiclassical wave function $\psi_\text{sc}(\mathbf r)$ in an intuitive, step-by-step approach.  (For formal reviews of the topic, see e.~g.\ Refs.~\onlinecite{Berry1972a,Maslov1981a,Delos1986a}.)  The basic idea is to assign a wave $\psi_\nu(\mathbf r)$ to every trajectory $\mathbf r_\nu(t)$ that leads from the source to the destination $\mathbf r$; the semiclassical wave function itself then is the sum of the individual wave amplitudes associated with the various paths $\mathbf r_\nu(t)$:
\begin{equation}
\label{eq:Semi1}
\psi_\text{sc}(\mathbf r) \,=\, \sum_{\nu = 1}^\infty  \psi_\nu(\mathbf r) \,=\, \sum_{j = \nu}^\infty |\psi_\nu(\mathbf r)| \, e^{i \Phi_\nu(\mathbf r)} \;.
\end{equation}
Note that $\psi_\text{sc}(\mathbf r)$ involves an infinite sum in our problem.  We now proceed to define the modulus $|\psi_\nu(\mathbf r)|$ and the phase $\Phi_\nu(\mathbf r)$ of each contribution.

The modulus $\psi_\nu(\mathbf r)$ is chosen so that the trajectory on its own contributes its classical weight $n_\nu(\mathbf r)$ (\ref{eq:Classical17}) to the particle density.  Since $|\psi(\mathbf r)|^2$ (\ref{eq:Quantum3}) yields the probability density, we set $|\psi_\nu(\mathbf r)| = \sqrt{n_\nu(\mathbf r)}$.  We note that the classical density diverges at the caustic set (the symmetry axis, and the nested turning surfaces displayed in Figure~\ref{fig:Caustics}).  Therefore, $\psi_\text{sc}(\mathbf r)$ will show unphysical behavior near the caustics.  Uniform approximations (see below) are available to remove the divergence, and replace it with a smooth transition of the wave function across the caustic.

For the phase $\Phi_\nu(\mathbf r)$, we adapt the De Broglie relation $\mathbf p = \hbar\mathbf k$ valid for a free particle to the accelerated motion of the charge in the magnetic field.  The classical momentum $\mathbf p_\nu(\mathbf r)$ then becomes position-dependent, and we postulate that the local change of the phase in space is again given by De Broglie's relation: $\bm\nabla \Phi_\nu(\mathbf r) = \mathbf k_\nu(\mathbf r) = \mathbf p_\nu(\mathbf r) / \hbar$.  Integrating along the trajectory from the source to the destination yields a phase difference:
\begin{equation}
\label{eq:Semi2}
\Delta\Phi_\nu^\text{dyn}(\mathbf r) \,=\, \frac 1\hbar \int_{\mathbf 0}^{\mathbf r} \mathbf p_\nu(\mathbf r') \cdot d\mathbf r' \,=\, \frac 1\hbar W_\nu(\mathbf r,\mathbf 0;E) \;,
\end{equation}
that is proportional to the classical Hamilton-Jacobi action functional $W_\nu(\mathbf r,\mathbf 0;E)$ \footnote{Since $\mathbf p$ is the canonical momentum, the wave fronts $\Phi = \text{const.}$ are not perpendicular to the trajectory field.}.  Using $\mathbf p = m\dot{\mathbf r} + q\mathbf A(\mathbf r)$, we rewrite Eq.~(\ref{eq:Semi2}) as a temporal integral, and insert the equation of motion (\ref{eq:Classical7}).  We integrate, switch to dimensionless coordinates $\hat\rho$ and $\hat z$, eliminate the emission angles, and find \footnote{Alternatively, using the time-dependent action (\ref{eq:Classical1}), $W_\nu(\mathbf r,\mathbf 0;E) = S(\mathbf r, t_\nu; \mathbf 0, 0) + E t_\nu $.}:
\begin{equation}
\label{eq:Semi3}
\Delta\Phi_\nu^\text{dyn}(\mathbf r) \,=\, \epsilon \left( \hat\rho^2 \cot\tau + \hat z^2/\tau + \tau \right) \;,
\end{equation}
where $\epsilon = E/(\hbar\omega_L)$ is the quantum mechanical energy parameter (\ref{eq:Quantum1}).  

Beside this ``dynamical'' contribution to the phase, one must also consider a discrete correction that traces back to the evolution of the particle density along the trajectory.  Eq.~(\ref{eq:Classical16}) shows that the classical density $n_\nu(\mathbf r)$ is a function of the determinant of the Jacobian $\mathcal J = \partial(\hat\rho,\hat z,\phi) / \partial(\theta_0,\phi_0,\tau)$, where a zero of $\det \mathcal J$ implies a singularity in the density.  For the semiclassical analysis, $n_\nu(\mathbf r)$ is an analytic function of $\det \mathcal J$, and changes its sign at each simple root of $\det\mathcal J$, i.~e., whenever the trajectory runs through a caustic.  Taking the square root of $n_\nu(\mathbf r)$ to find the modulus of the semiclassical wave function, each sign change in $n_\nu(\mathbf r)$ translates into an additional factor $-i$, which we include into the phase $\Phi_\nu$ as a discrete shift of $- \frac{\pi}2$ \footnote{In one-dimensional problems, this scheme yields the well-known WKB connection formula.}.  The {\sl Maslov index} $\mu_\nu$ denotes the number of sign changes of $\det \mathcal J$ along a trajectory $\mathbf r_\nu(t)$, so the total phase accumulated from $\mathbf 0$ to $\mathbf r$ is:
\begin{equation}
\label{eq:Semi4}
\Delta\Phi_\nu(\mathbf r) \,=\, \Delta\Phi_\nu^\text{dyn}(\mathbf r) - \frac \pi2 \mu_\nu \;.
\end{equation}
The Maslov index $\mu_\nu$ can be read off Eq.~(\ref{eq:Classical17}) in a straightforward manner.  Whenever the charge has completed a cyclotron orbit and returns to the symmetry axis, $\sin\tau = 0$ holds, and $\det\mathcal J$ changes sign.  In addition, within each cyclotron interval, $d{\hat\epsilon}/d\tau$ drops monotonically from $+\infty$ to $-\infty$, implying one more simple root of $\det\mathcal J$ which corresponds to the turning point of the path on the fold caustics.  Hence, with every completed cyclotron orbit, the Maslov index grows by two.  In the final arc of the orbit, when the trajectory reaches its destination, the path arrives either before touching the caustic, or afterwards, corresponding to the ``fast'' and ``slow'' solutions in this cyclotron interval shown in Figure~\ref{fig:Epsilon}, respectively.  (Note that these solutions differ in the sign of $d{\hat\epsilon}/d\tau$.)  Hence, the Maslov index $\mu_\nu$ for a solution in the $\nu$th cyclotron interval $\nu\pi < \tau < (\nu+1)\pi$ is:
\begin{equation}
\label{eq:Semi5}
\mu_\nu \,=\, 
\begin{cases}
2\nu & (\text{``fast'' path, }d{\hat\epsilon}/d\tau > 0) \;, \\
2\nu +1 & (\text{``slow'' path, }d{\hat\epsilon}/d\tau < 0) \;.
\end{cases}
\end{equation}
Note that the semiclassical wave function $\psi_\text{sc}(\mathbf r)$ can be expressed as a function of the times of flight $\tau_\nu$ of the trajectories.

In order to compare the quantum result to the semiclassical approximation quantitatively, we finally need to fix the previously unspecified emission rate $J$ of the classical source in Eq.~(\ref{eq:Classical15}).  For this purpose, we identify $J$ with the quantum mechanical current $J_\text{free}(E) = mk / (\pi\hbar^3)$ (\ref{eq:Appendix8}) emitted by a unit source of {\sl free} particles in the absence of the magnetic field (see Appendix).

To improve the semiclasssical approximation, we also include tunneling trajectories into our calculations.  For these, conjugate complex solution pairs for the times of flight $\tau_\nu$ are obtained.  Their semiclassical contributions follow from complex continuation of the expression for the dynamical phase (\ref{eq:Semi3}); only the physically acceptable solution which leads to an exponentially decaying wave function is included with the sum (\ref{eq:Semi1}).

\subsection{Convergence properties}

Recall that the number of classical trajectories connecting source and destination in the magnetic field environment is infinite, and that the concept of a stationary emitter is ill-defined in a purely classical description.  In the semiclassical picture, the wave function $\psi_\text{sc}(\mathbf r)$ therefore becomes an infinite sum (\ref{eq:Semi1}), and it is of interest to study its convergence as a function of the particle energy $\epsilon$.  For this purpose, it suffices to consider the behavior of the individual waves $\psi_\nu$ making up the series in the asymptotic limit $\nu \rightarrow \infty$.

In the long-time limit, the classical density $n_\nu(\mathbf r)$ (\ref{eq:Classical18}) drops inversely with the number of cyclotron orbits $\nu$.  For the complex series (\ref{eq:Semi1}), we now additionally inquire into the limiting behavior of the phase $\Delta\Phi_\nu$ (\ref{eq:Semi4}) for large $\nu$.  Since $\sin\tau \sim \hat\rho$ holds in the long-time limit, we find $\cot\tau \sim \pm \sqrt{1- {\hat\rho}^2} /{\hat\rho}$, so the phases for the ``fast'' and ``slow'' trajectories asymptotically approach the values:
\begin{equation}
\label{eq:Semi6}
\Delta\Phi_\nu \,\sim\, 
\begin{cases}
\epsilon \left( \nu\pi + \arcsin\hat\rho + \hat\rho\sqrt{1-{\hat\rho}^2} \right) - \nu\pi & (\text{``fast'' path}) \;, \\
\epsilon \left( (\nu +1) \pi - \arcsin\hat\rho - \hat\rho\sqrt{1-{\hat\rho}^2} \right) - \left(\nu +\frac12 \right)\pi & (\text{``slow'' path}) \;,
\end{cases}
\end{equation}
as $\nu \rightarrow\infty$.  (Note the second contribution arising from the growing Maslov index $\mu_\nu$ (\ref{eq:Semi5}).)  In either case, as we increment the number of cyclotron orbits $\nu \rightarrow \nu+1$, the phase asymptotically increases by an overall amount $\pi(\epsilon - 1)$.  Thus, the semiclassical series has the asymptotic form:
\begin{equation}
\label{eq:Semi7}
\psi_\text{sc}(\mathbf r) \,\sim\, \left( C_\text{fast}(\hat\rho) + C_\text{slow}(\hat\rho) \right) \sum_{\nu}  
\frac{e^{i\pi(\epsilon - 1)\nu}}{\sqrt \nu} \;,
\end{equation}
where the prefactor is a function of the lateral distance $\hat\rho$ only, and can be read off Eqs.~(\ref{eq:Classical18}) and (\ref{eq:Semi6}).  

In the mathematical literature, the complex sum in (\ref{eq:Semi7}) is known as a {\sl periodic zeta function}~$F[\frac12 (\epsilon -1), \frac12]$ \cite{Olver2010b}.  Unless the phase increase in the exponent is a multiple of $2\pi$, the sum is alternating, and conditionally convergent.  Otherwise, the sum is real and divergent, growing with the square root of the summation limit.  This happens whenever $\epsilon = \epsilon_l = 2l+1$ is an odd integer.  Hence, the semiclasssical approximation reproduces a key property of the quantum solution: Unlike the classical density, $\psi_\text{sc}(\mathbf r)$ is well-defined, unless the energy of the particles coincides with one of the Landau levels in the magnetic field.

In practice, convergence of the series (\ref{eq:Semi7}) is slow, in particular in the vicinity of the Landau level thresholds where $\epsilon$ almost matches an odd integer value.  In numerical simulations, we found it necessary to adjust the number of trajectories included in the summation to achieve good agreement with the quantum solution.  We experimented with a number of sophisticated schemes to accelerate convergence of the semiclassical series, but found no consistent superior performance compared to a ``hard cut-off'' in the summation.  We therefore adopted this simple method in our simulations below. 

\subsection{Semiclassical density and current}

Once the semiclassical wave function $\Psi_\text{sc}(\mathbf r)$ (\ref{eq:Semi1}) is established, the analysis proceeds in the same vein as in the quantum case.  As in Eq.~(\ref{eq:Quantum3}), the approximation to the particle density $n_\text{sc}(\mathbf r) = |\psi_\text{sc}(\mathbf r)|^2$ is given by the absolute square of the wave function, and now becomes a sum over all pairs of trajectories:
\begin{equation}
\label{eq:Semi8}
n_\text{sc}(\mathbf r) \,=\, \sum_{\alpha,\beta = 1}^\infty  \sqrt{n_\alpha(\mathbf r) n_\beta(\mathbf r)} \, e^{i [\Phi_\alpha(\mathbf r) - \Phi_\beta(\mathbf r)]} \;.
\end{equation}
It is worth noting that this sum, like the semiclassical wave function itself, is conditionally convergent.  As a result, it cannot be reordered into a classical density $n_\text{cl}(\mathbf r)$, given by the diagonal terms with $\alpha=\beta$, and ``interference terms'' with $\alpha\neq \beta$, as the classical density diverges (see Section~\ref{sec:Classical}).

For the current density $\mathbf j_\text{sc}(\mathbf r)$, we start from the quantum expression (\ref{eq:Quantum4}), and replace the operator $\mathbf p - q\mathbf A(\mathbf r)$ with its classical counterpart, the kinematic momentum  $m\dot{\mathbf r}(t)$ \footnote{Note that the gradient of the semiclassical wave function $\psi_\text{sc}(\mathbf r)$ yields the classical velocity $\mathbf v$ via the derivative of the dynamical phase $\Delta\Phi_\nu(\mathbf r)$, and an additional contribution from the change in classical density $\nabla \sqrt{n_\nu(\mathbf r)}$.  We neglect the latter term for two reasons: First, it is of order $\hbar$ and thus vanishes in the classical limit, and second, it becomes largest at the caustics where the semiclassical method is known to fail anyway.}.  We obtain again a sum over pairs of trajectories, now weighted with the mean particle velocity $\mathbf v(\mathbf r)$ at the destination point:
\begin{equation}
\label{eq:Semi9}
\mathbf j_\text{sc}(\mathbf r) \,=\, \frac12 \sum_{\alpha,\beta = 1}^\infty  \sqrt{n_\alpha(\mathbf r) n_\beta(\mathbf r)} \left[\mathbf v_\alpha(\mathbf r) + \mathbf v_\beta(\mathbf r)\right] e^{i [\Phi_\alpha(\mathbf r) - \Phi_\beta(\mathbf r)]} \;.
\end{equation}
This is suggestive of the classical relation $\mathbf j = n\mathbf v$, but we note again that the sum (\ref{eq:Semi9}), due to its conditionally convergent nature, cannot be reordered at will.  In fact, the current flowing from a source in a magnetic field features rather counterintuitive behavior, as discussed in Section~\ref{sec:Results}.

\subsection{Uniform approximation}

Because the classical density $n_\text{cl}(\mathbf r)$ becomes singular there, the ``primitive'' semiclassical approximation (\ref{eq:Semi1}) to the wave function is bound to fail near the caustics.  Still, it is possible to find higher-level {\sl uniform approximations}, based on classical trajectories, that correct the divergent behavior.  The idea behind these approximations is that caustic points share generic types of divergence (``catastrophes'') \cite{Thom1975a,Berry1976a,Poston1978a,Berry1981b,Nye1999a}, and therefore a wave solution valid in the vicinity of any such point provides a template for the solution at all related points.  For the simplest catastrophe, the fold-type caustic, the prototype solution \cite{Berry1976a,Schulman1981a,Nye1999a} is an Airy function $\mathop{\text{Ai}}(u)$ \cite{Olver2010a}, the wave solution for a quantum particle ``turning around'' under a constant force in one dimension \footnote{For an elementary discussion of the problem, see e.~g., D.~J.~Griffiths, {\sl Introduction to Quantum Mechanics} (2nd ed.), Pearson Prentice Hall (Upper Saddle River, NJ) (2005), p.~325--335.}. 

Fold caustics are associated with pairs of trajectories coalescing and disappearing, which occurs in the magnetic field problem at the nested onion-shaped turning surfaces parametrized by Eq.~(\ref{eq:Classical14}), and displayed in Fig.~\ref{fig:Caustics}.  Hence, in the vicinity of these points, the contribution of the affected pair of trajectories to the semiclassical wave function (\ref{eq:Semi1}) should be replaced by an Airy function, with an argument appropriately matched to the change in their dynamical phase $\Delta\Phi_\nu^{\text{dyn}}$ (\ref{eq:Semi2}).  In our simulations, we adapted a technique developed in Ref.~\onlinecite{Bracher2006b}, and assigned a combination of an Airy function and its derivative to each pair of ``fast'' and ``slow'' classical paths within a cyclotron period.  Thus, the uniform approximation used is an infinite sum of Airy functions.

The expansion in Airy function yields excellent results near the turning surfaces, but is not suitable for the second type of caustics encountered in our problem, the ``focal line'' $\hat\rho = 0$.  Bundles of trajectories periodically converge upon this line under all angles $\phi$, and their interference yields a Bessel function of order zero $J_0(u)$ as an amplitude profile.  (The cylindrical cusps that form the joints between turning surfaces and focal lines need to be considered separately \cite{Peters1997b}.  Neither focal lines nor cylindrical cusps are ``generic'' catastrophes in the mathematical sense, but are commonly encountered in systems with cylindrical symmetry.)  Hence, the wave function at small $\hat\rho \rightarrow 0$ is more appropriately described by a superposition of an infinite number of such Bessel functions.  While interesting in its own right, we did not attempt to model the wave function near the symmetry axis in this way.

\section{Results}
\label{sec:Results}

In this section, we will present the findings from numerical studies for a representative set of values for the energy parameter $\epsilon$ (\ref{eq:Quantum1}).  Considering the simplicity of the setup, a surprisingly rich set of features is found in these simulations, some of which defy easy explanation.  Another objective of our studies is to assess the performance of the semiclassical and uniform approximations in comparison to the exact quantum results.  We have pointed out that the semiclassical method operates here under challenging conditions, with an infinite number of classical paths present, wave functions represented by conditionally convergent series, and the classical counterpart of the problem being altogether ill-defined.  However, we find that in all cases studied, the semiclassical technique produced reliable results, with an accuracy only limited by the time and depth allotted to the computations.

\subsection{Density profiles between Landau levels}

For our first case study, we choose $\epsilon = 50$ as a value for the energy parameter.  This places the electron energy right between the 24th and the 25th Landau level, whose thresholds are located at 49 $\hbar\omega_L$ and 51 $\hbar\omega_L$, respectively.  Hence, 24 open scattering channels contribute to the electronic current, while the quantum wave function $G(\mathbf r, \mathbf 0;E)$ (\ref{eq:Quantum2}) has additional evanescent components that modify the density and current profile near the source.  Since we pick a value of $\epsilon$ removed from the thresholds $\epsilon_l = 2l+1$, the various scattering channels have comparable contributions, none clearly dominating the others.  Hence, this example stands for a ``regular'' situation, with results ``typical'' for most values of $\epsilon$.

\begin{figure}[p]
\includegraphics[width=0.75\textwidth]{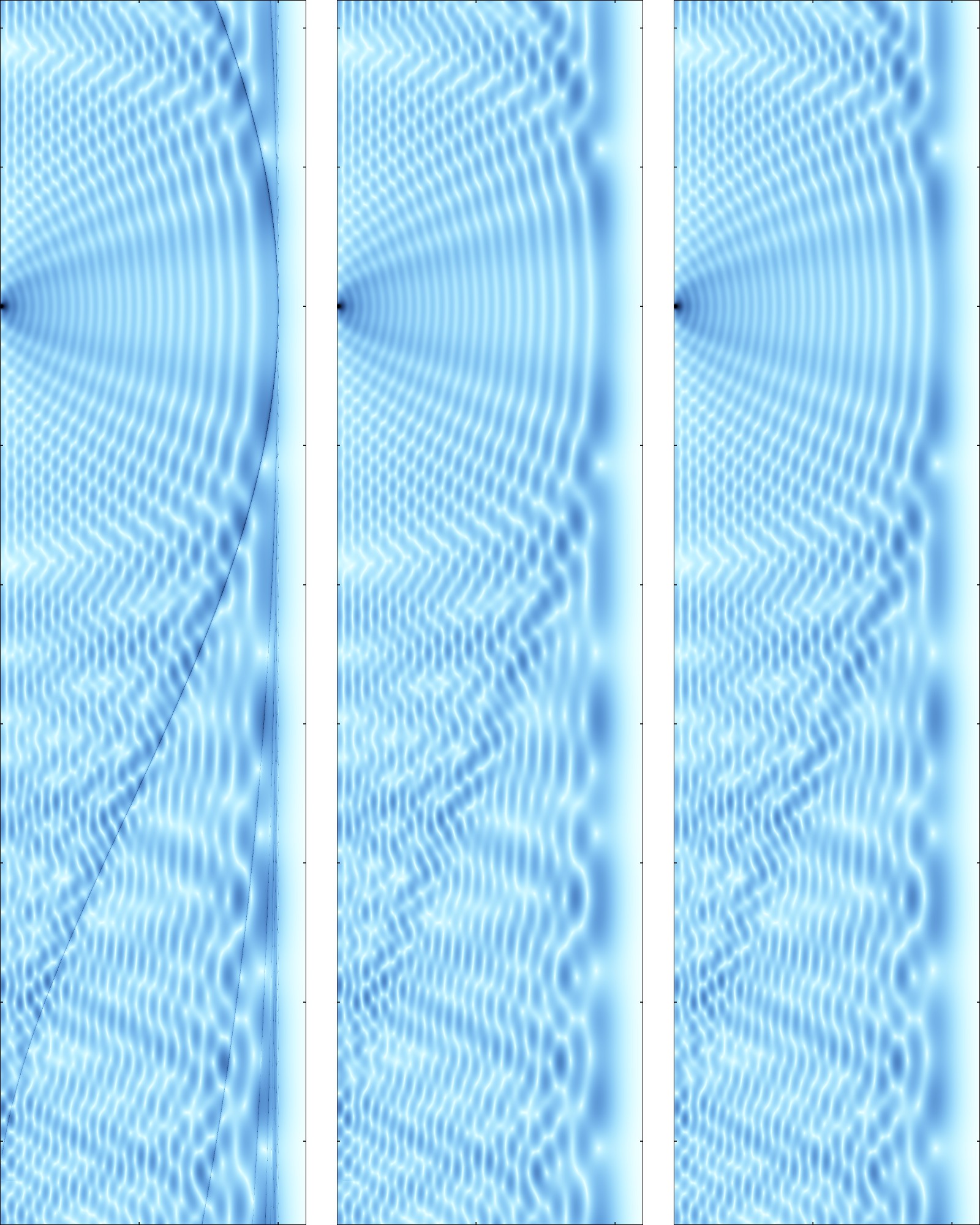}
\caption{\label{eps50:densitymap}
Integrated charge density $2\pi\hat\rho n(\hat\rho, \hat z)$ emitted by an isotropic point source in a magnetic field (vertical direction), along a radial cut in the $\hat\rho - \hat z$ plane, for dimensionless energy $\epsilon = 50$.  Left panel:  Primitive semiclassical calculation (including tunneling trajectories).  Center panel:  Uniform approximation based on Airy functions.  Right panel:  Exact quantum result, evaluated using Eq.~(\ref{eq:Quantum2}).  
Up to 500 trajectories have been summed for the approximations.  The dark ``streaks'' tracing the location of the caustics in the left panel are caused by the divergence of the semiclassical approximation there, which otherwise agrees with the quantum simulation (right panel).  The uniform (center) and quantum results (right) are visually indistinguishable. ---  The images cover the range $0 \leq \hat\rho \leq 1.1$ in horizontal direction, and $-1.1 \leq \hat z \leq 3.3$ in vertical direction.  The source is conspicuous at the upper left edge of each image.  Dark spots correspond to high density.  All units dimensionless.}
\end{figure}

\begin{figure}[t]
\includegraphics[width=0.75\textwidth]{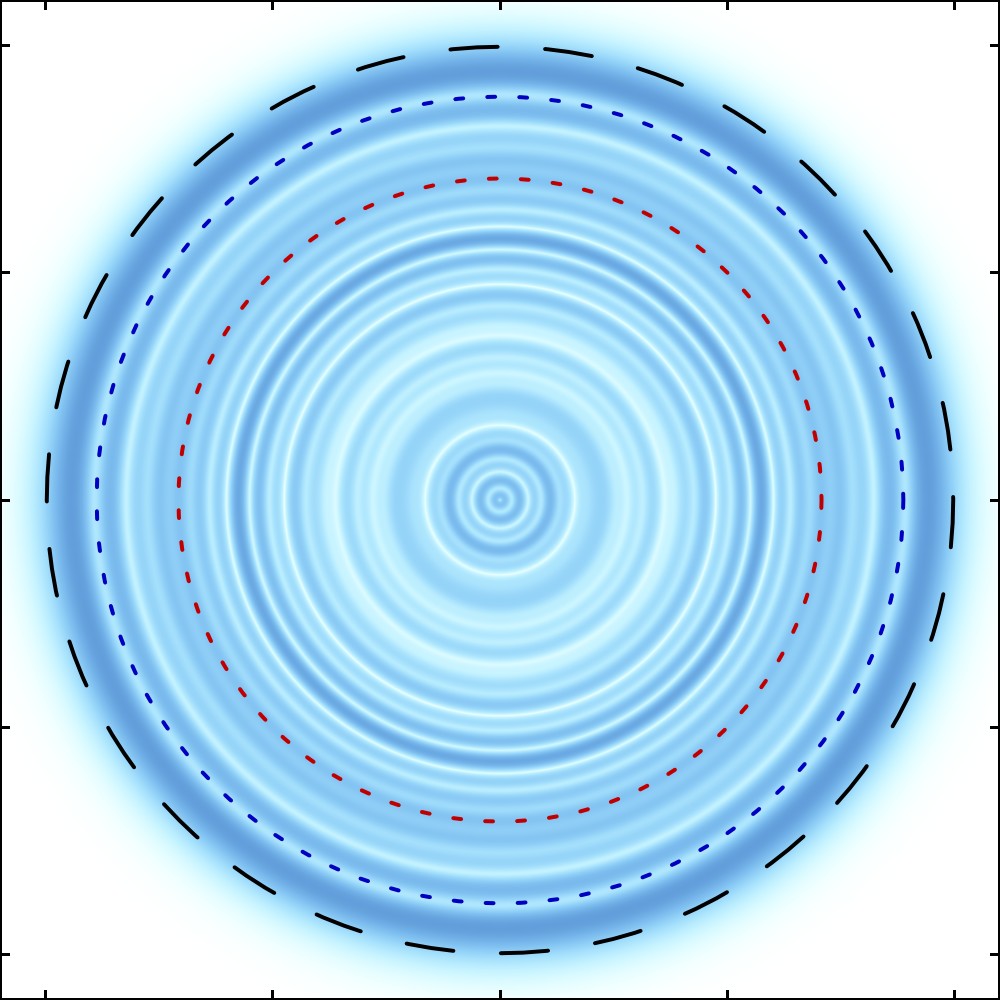}
\caption{\label{eps50:fringemap}
Integrated charge density along a slice perpendicular to the magnetic field for energy $\epsilon = 50$, at a scaled dimensionless distance $\hat z = 3.3$, corresponding to the bottom edge of the radial density maps in Figure~\ref{eps50:densitymap}.  The frame extends 1.1 units from the center in $\hat x$- and $\hat y$-direction.  Due to cylindrical symmetry, the quantum simulation shows circular interference fringes. For comparison, the concentric circles indicate the intersections with the 2nd and 3rd turning surface (inner and outer short dashed lines), and the boundary of classical motion $\hat\rho = 1$ (long dashed line).}
\end{figure}

We first examine the charge density $n(\mathbf r)$ generated by the source.  In Figure~\ref{eps50:densitymap}, we plot color-coded maps of the density in the $\hat\rho-\hat z$ plane, i.~e., a cut through the three-dimensional distribution parallel to the magnetic field axis that contains the source.  The figure shows the results of a semiclassical calculation (left panel) using Eq.~(\ref{eq:Semi8}), a simulation using the uniform approximation detailed in the previous section, based on the same trajectories (center panel), and the exact quantum distribution (\ref{eq:Quantum3}) (right panel).  For better display, we plotted the ``integrated'' density $2\pi\hat\rho n(\hat\rho, \hat z)$, summing up the contributions for various azimuthal angles $\phi$.  Dark (blue) patches in the images correspond to high density, white to low density.  Owing to the cylidrical symmetry inherent in the problem, a section perpendicular to the magnetic field axis yields an interference pattern of concentric fringes that does not reveal any additional information.  Figure~\ref{eps50:fringemap} shows such a density distribution, where the cut is taken along the bottom edge of Figure~\ref{eps50:densitymap}.
%
%
%
%

The semiclassical calculations in the panels were performed with an arbitrary time-of-flight cutoff at $\tau = 250\pi$, corresponding to at most 500 classical orbits, to keep the time of calculation reasonable \footnote{In the most time-consuming simulations, generation of a 4 megapixel image using the uniform approximation required about one hour on a standard PC.  By contrast, evaluating the quantum solution for the same image takes only seconds.}.  Nevertheless, the semiclassical methods reproduce the quantum density distribution very well.  As expected, the ``primitive'' semiclassical computation fails at the location of the caustics, where the calculated density diverges, leading to linear ``streaks'' in the image.  The uniform approximation corrects this unphysical behavior, and provides a density map that almost perfectly matches the quantum result.  Note that the quantum image clearly displays a strong increase in density along the location of the first ``onion-like'' caustic (see Figure~\ref{fig:Caustics}), and a set of interference fringes running parallel to it, despite the fact that the quantum calculation never invokes the concept of trajectories.  This illustrates the power of the semiclassical method to explain features of the quantum solution.

%
\begin{figure}[t]
\includegraphics[width=0.75\textwidth]{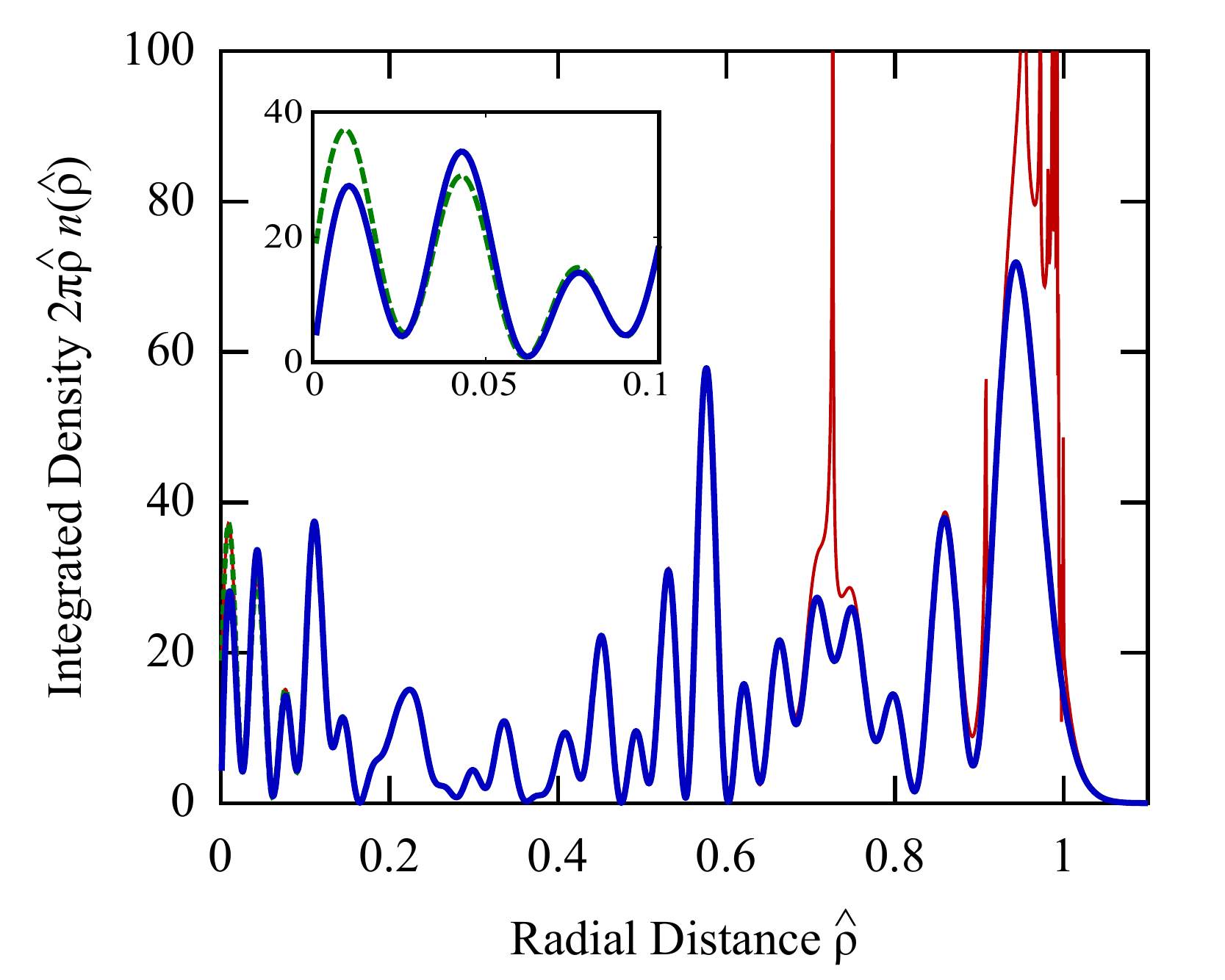}
\caption{\label{eps50:densityprofile}
Radial profiles of the integrated charge density $2\pi\hat\rho n(\hat\rho)$ at a vertical distance $\hat z = 3.3$ from the source, for an energy $\epsilon = 50$, using units of $[mk / (4\pi\epsilon\hbar^2)]^2$ (\ref{eq:Appendix6}).  The plot displays the primitive semiclassical approximation (\ref{eq:Semi8}) (thin red curve), uniform approximation (dashed green curve), and exact quantum result (\ref{eq:Quantum3}) (bold blue curve), as shown in Figure~\ref{eps50:densitymap} (bottom edge of maps); the blue curve correponds directly to the density cross section displayed in Figure~\ref{eps50:fringemap}.  Up to $5\cdot10^4$ trajectories have been included in the semiclassical calculation.  All three methods yield good agreement, except that the semiclassical result diverges at the intersections with the turning surfaces, a failure that does not affect the uniform approximation.  Both methods based on classical orbits are unable to reproduce the quantum density profile near the symmetry axis (inset detail plot).}
\end{figure}
For a quantitative assessment of the semiclassical approximation, we also calculated a radial profile of the probability density, taken at $\hat z = 3.3$, corresponding to the bottom edge of Figure~\ref{eps50:densitymap}, and representing the circular fringes in Figure~\ref{eps50:fringemap}.  Here, we increased the cutoff time to $\tau = 25000\pi$, and thus added the contributions of up to 50,000 classical orbits for the semiclassical and uniform approximations.  The resulting densities in ``natural units'' $[mk / (4\pi\epsilon\hbar^2)]^2$ derived from the free-particle Green function (see Eq.~(\ref{eq:Appendix6}) in the Appendix) are plotted in Figure~\ref{eps50:densityprofile}.  Both approximations quantitatively coincide with the exact result sufficiently far from the caustics.  Whereas the primitive semiclassical method shows divergence at the intersections with the turning surfaces for $\hat\rho > 0$, the uniform approximation is not affected and provides results virtually indistinguishable from the quantum calculation.  Both methods deviate from the exact result in the vicinity of the symmetry axis $\hat\rho \rightarrow 0$.  This is not surprising, as $\hat\rho = 0$ is itself part of the caustics, and a uniform expansion into Airy functions is not appropriate for the focal line structure there.

\begin{figure}[p]
\includegraphics[width=0.75\textwidth]{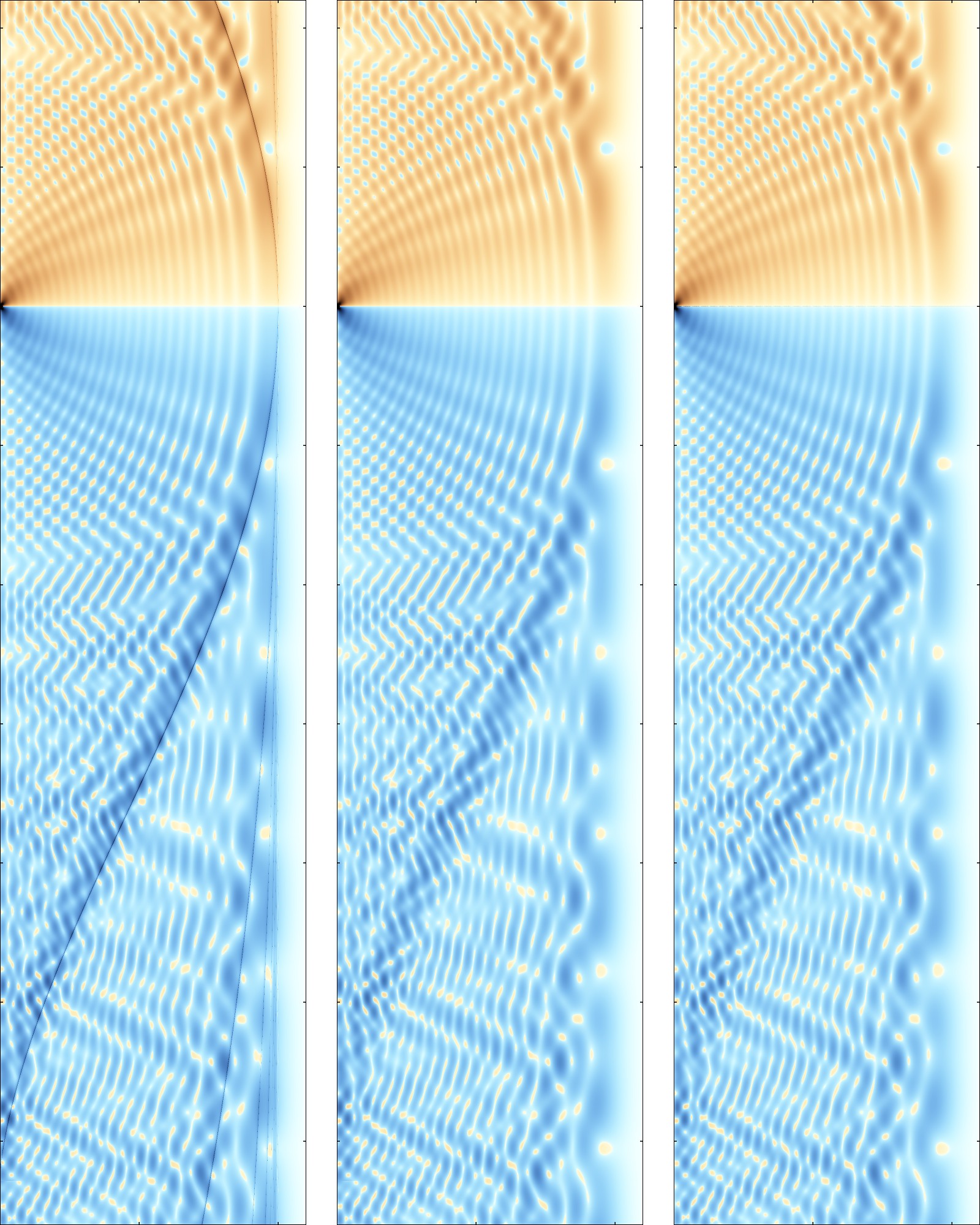}
\caption{\label{eps50:currentmap}
(Color online) Integrated current density in field direction $2\pi\hat\rho j_z(\hat\rho, \hat z)$ in the $\hat\rho - \hat z$ plane.  Left panel:  Primitive semiclassical calculation.  Center panel:  Uniform approximation based on Airy functions.  Right panel:  Exact quantum result, evaluated using Eq.~(\ref{eq:Quantum5}).  All parameters are as in Figure~\ref{eps50:densitymap}.  Dark spots correspond to high current density; blue (red) indicates downward (upward) orientation.}
\end{figure}

\begin{figure}[t]
\includegraphics[width=0.75\textwidth]{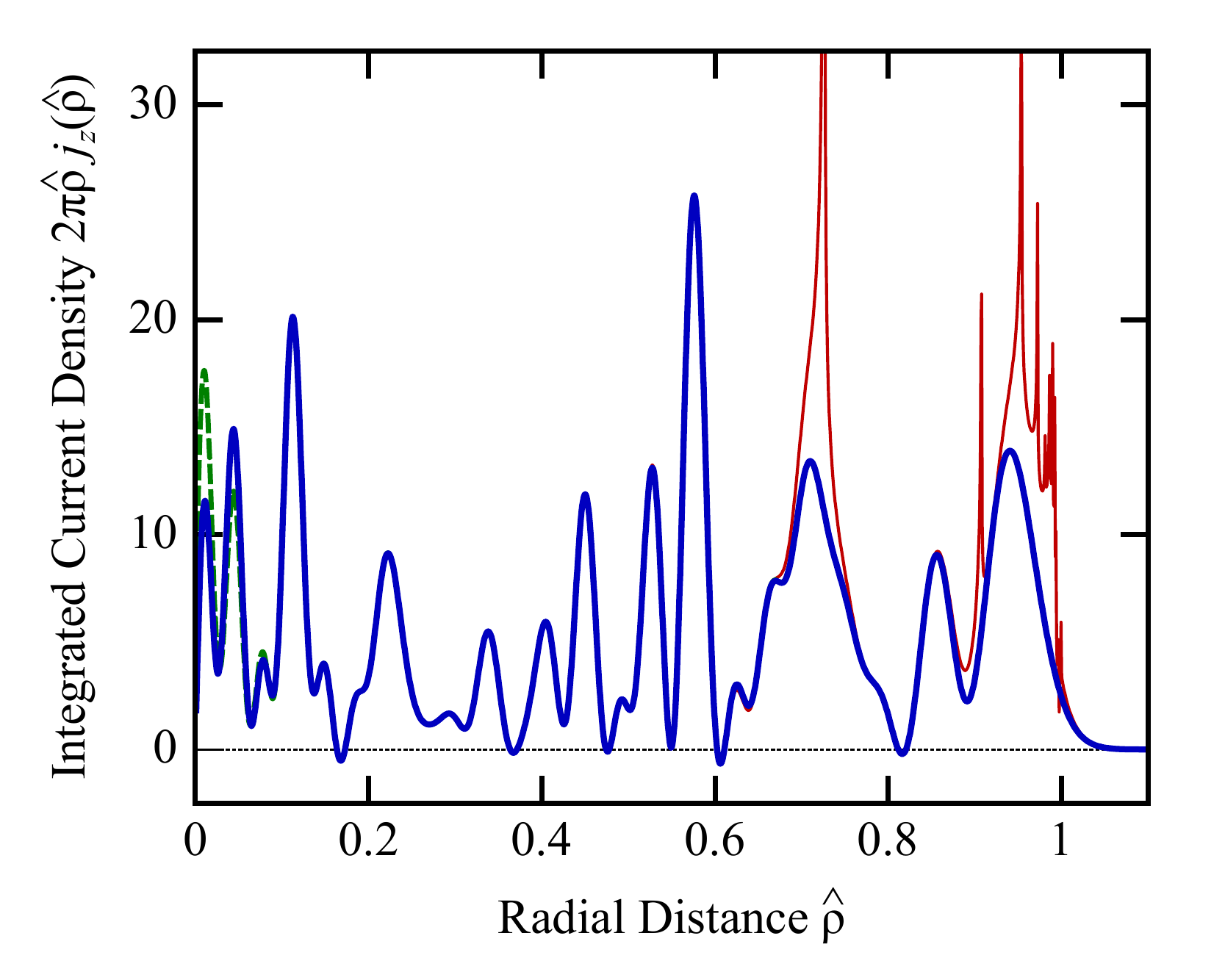}
\caption{\label{eps50:currentprofile}
(Color online) Radial profiles of the integrated current density $2\pi\hat\rho j_z(\hat\rho)$ for a distance $\hat z = 3.3$ and energy $\epsilon = 50$, in units of $mk^3 / (16\pi^2\epsilon^2\hbar^3)$ (\ref{eq:Appendix7}).  As in Figure~\ref{eps50:densityprofile}, the primitive semiclassical approximation (\ref{eq:Semi9}) (thin red curve) displays singularities at the intersection with caustics.  The uniform approximation (dashed green curve) and exact quantum result (\ref{eq:Quantum3}) (bold blue curve) are in excellent agreement, except in the vicinity of the focal line $\hat\rho = 0$.  Note that all three methods indicate a reversal of the current at the center of some interference minima.}
\end{figure}

In a second step, we repeat a similar set of calculations for the current density $\mathbf j(\mathbf r)$.  We first concentrate on the current density component $j_z(\mathbf r)$ aligned with the field direction, and perform again a comparison of the semiclassical and uniform approximations (\ref{eq:Semi9}) with the quantum result (\ref{eq:Quantum5}).  A current map, using the same parameters as the density map (Figure~\ref{eps50:densitymap}), is displayed in Figure~\ref{eps50:currentmap}.  We find again good agreement of the three approaches, except near the caustics, where the semiclassical approximation fails, as expected.  While differing in detail, the current distribution shares the same qualitative features observed in the charge density distribution, including enhancement and interference along the prominent onion-type caustic.  The most striking difference between the two maps is that the flow of particles is reversed in the upper part of the image (upward currents are encoded in red, downward currents in blue), simply confirming the expectation that the particles stream away from the source in either direction.  Again, a more quantitative comparison is undertaken in the current profile displayed in Figure~\ref{eps50:currentprofile}, corresponding to the bottom edge of the current maps in Figure~\ref{eps50:currentmap}, computed using the same conditions as in Figure~\ref{eps50:densityprofile}.  The plot (which displays the current density in ``natural'' units of $mk^3 / (16\pi^2\epsilon^2\hbar^3)$ (\ref{eq:Appendix7}), extracted from an analysis of a free-particle source) confirms the observations about the convergence of the three calculations we made before.  Careful scrutiny reveals an interesting detail:  In the center of interference minima, $j_z(\hat\rho)$ drops below zero, indicating reversal of the flux in the upward direction.  All three methods of computation agree on this counterintuitive ``backflow'' phenomenon.
%
%

%

\begin{figure}[p]
\includegraphics[width=0.75\textwidth]{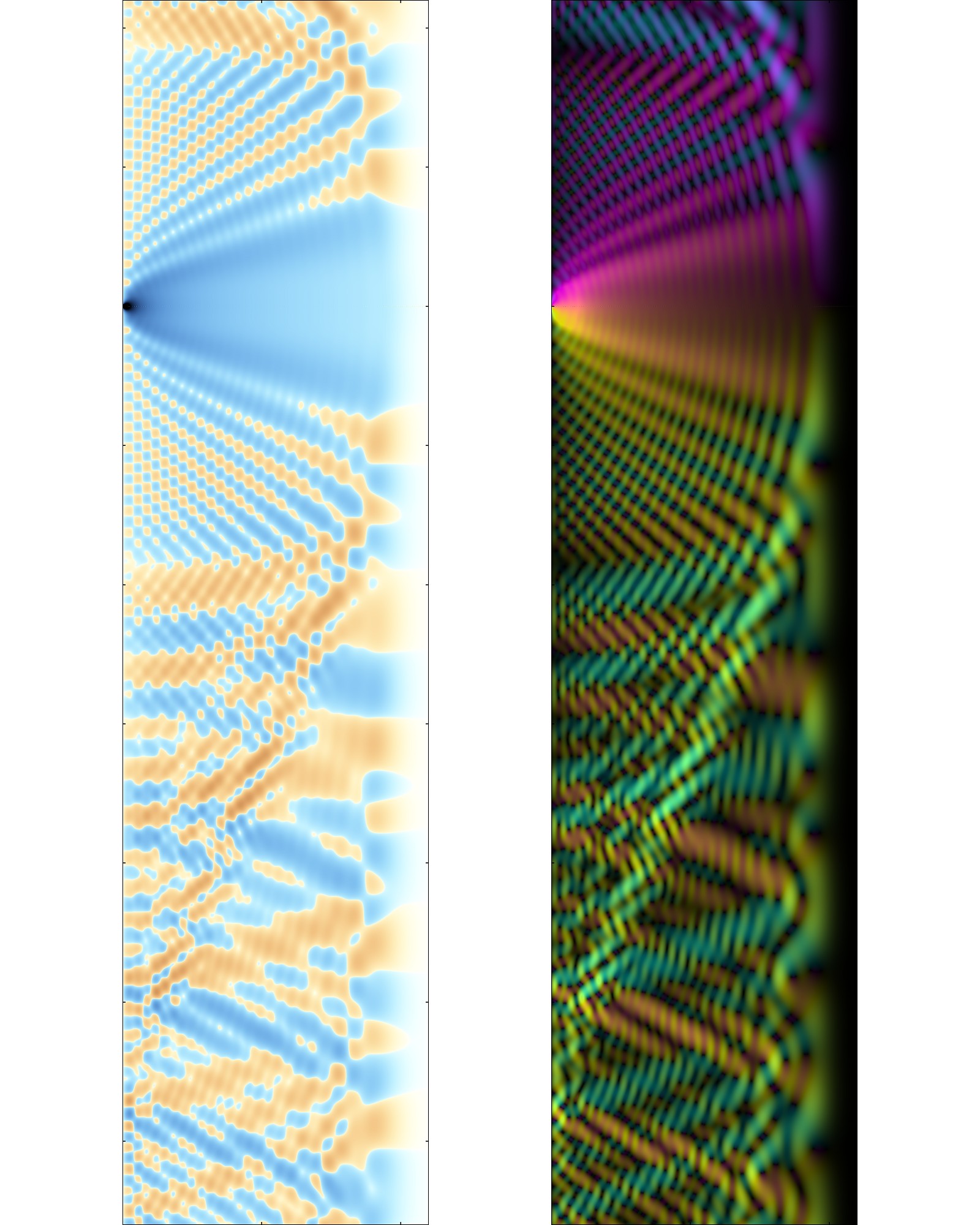}
\caption{\label{eps50:currentfield}
Maps of the integrated current density vector $2\pi\hat\rho \mathbf j(\hat\rho, \hat z)$ in the $\hat\rho - \hat z$ plane for energy $\epsilon = 50$.  Left panel:  Radial component $j_\rho(\hat\rho, \hat z)$ of the current density.  Blue indicates outward flow from the symmetry axis, red inward flow; dark spots indicate high current. --- Right panel:  Color-coded map of the current flows in the $\hat\rho - \hat z$ plane.  Brightness corresponds to current intensity, hue to the direction of the current (red -- to the right, green -- to the lower left, blue -- to the upper left).  Charge transport occurs both along the caustics, and in peculiar bands (orange and teal pattern).  (Parameters as in Figure~\ref{eps50:densitymap}.)}
\end{figure}

Finally, we also performed quantum calculations for the radial current component $j_\rho(\mathbf r)$ (\ref{eq:Quantum5}), displayed as a map in Figure~\ref{eps50:currentfield} (left panel).  Here, the current is seen to initially stream away from the axis $\hat\rho = 0$ in the vicinity of the source, but ultimately a complicated pattern of alternating inward flows (red) and outward flows (blue) ensues that is difficult to explain from a classical point of view.  To illustrate the streaming pattern of the charge, we have combined the radial and parallel components of $\mathbf j(\mathbf r)$ into a ``quantum flow map'' (right panel in Figure~\ref{eps50:currentfield}).  Here, brightness corresponds to the magnitude of the current, whereas the color space indicates the direction of the current probability vector in the $\hat\rho - \hat z$ plane, with red indicating flux to the right, blue to the upper left, and green to the lower left.  Transport occurs parallel to the caustic surface, but also along distinct paths of unexplained etiology that criss-cross the classically allowed sector from the symmetry axis ($\hat\rho = 0$) to the outer limits of motion ($\hat\rho = 1$), visible as orange and teal bands in the image.  (No plot of the azimuthal current density $j_\phi(\mathbf r)$ is provided, as it is simply proportional to the particle density $n(\mathbf r)$ (\ref{eq:Quantum6}).  Classically, the whole distribution rotates uniformly with frequency $\omega_L$.)

\subsection{Density profiles near threshold}

We now slightly increase the particle energy to $\epsilon = 51.01$, a value just above the threshold of the 25th Landau level at $\epsilon_{25} = 51$.  Whereas all other terms in the series solution for the energy Green function $G(\mathbf r, \mathbf 0;E)$ (\ref{eq:Quantum2}) undergo only gradual change, the newly opened scattering channel adds an outsized contribution to this sum, as the relative weight of a given Landau level $l$ depends inversely on the energy $\epsilon - 2l - 1$ available for the motion in field direction.  The new emission mode flushes the environment with slowly drifting particles, and we expect that the shape of the electronic density distribution $n(\mathbf r)$ resembles the density profile of the dominant eigenstate $\psi_{l0}(\hat\rho)$ (\ref{eq:Appendix12}) with index $l = 25$ of the charge in the magnetic field (see Appendix).  The influence of the new scattering channel on the current distribution $\mathbf j(\mathbf r)$ (\ref{eq:Quantum4}) is less obvious, as the abundance of particles in this channel and their slow drift velocity have opposite effects on $\mathbf j(\mathbf r)$.
%
%
\begin{figure}[p]
\includegraphics[width=0.75\textwidth]{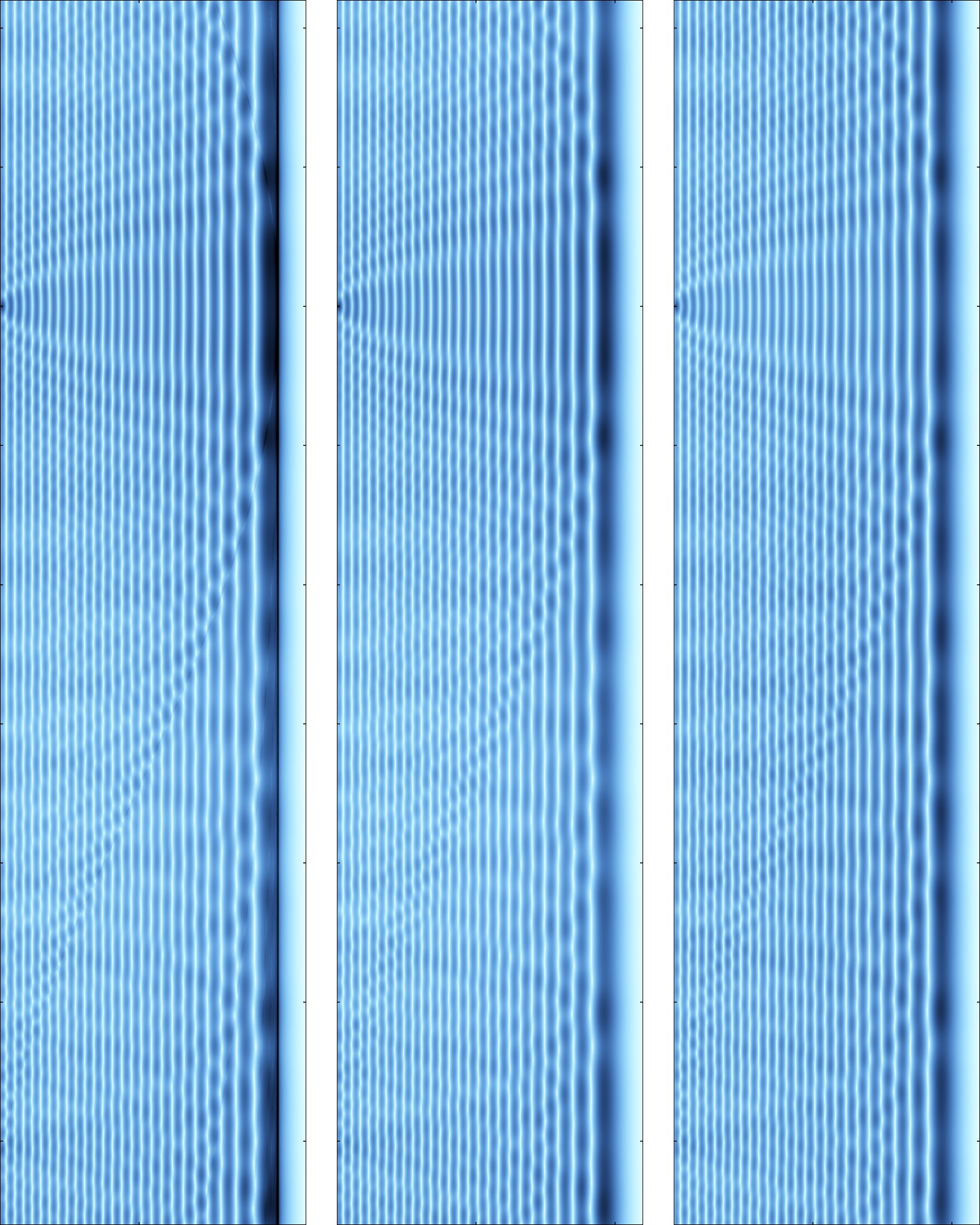}
\caption{\label{eps51:densitymap}
Integrated charge density $2\pi\hat\rho n(\hat\rho, \hat z)$ in the $\hat\rho - \hat z$ plane, for energy $\epsilon = 51.01$.  Shown:  Semiclassical (left panel) and uniform (center panel) approximations, and exact quantum result (right panel).  Other parameters as in Figure~\ref{eps50:densitymap}.  The banded appearance is due to the dominance of the scattering channel associated with the Landau level $l =25$.}
\end{figure}
%
%
\begin{figure}[t]
\includegraphics[width=0.75\textwidth]{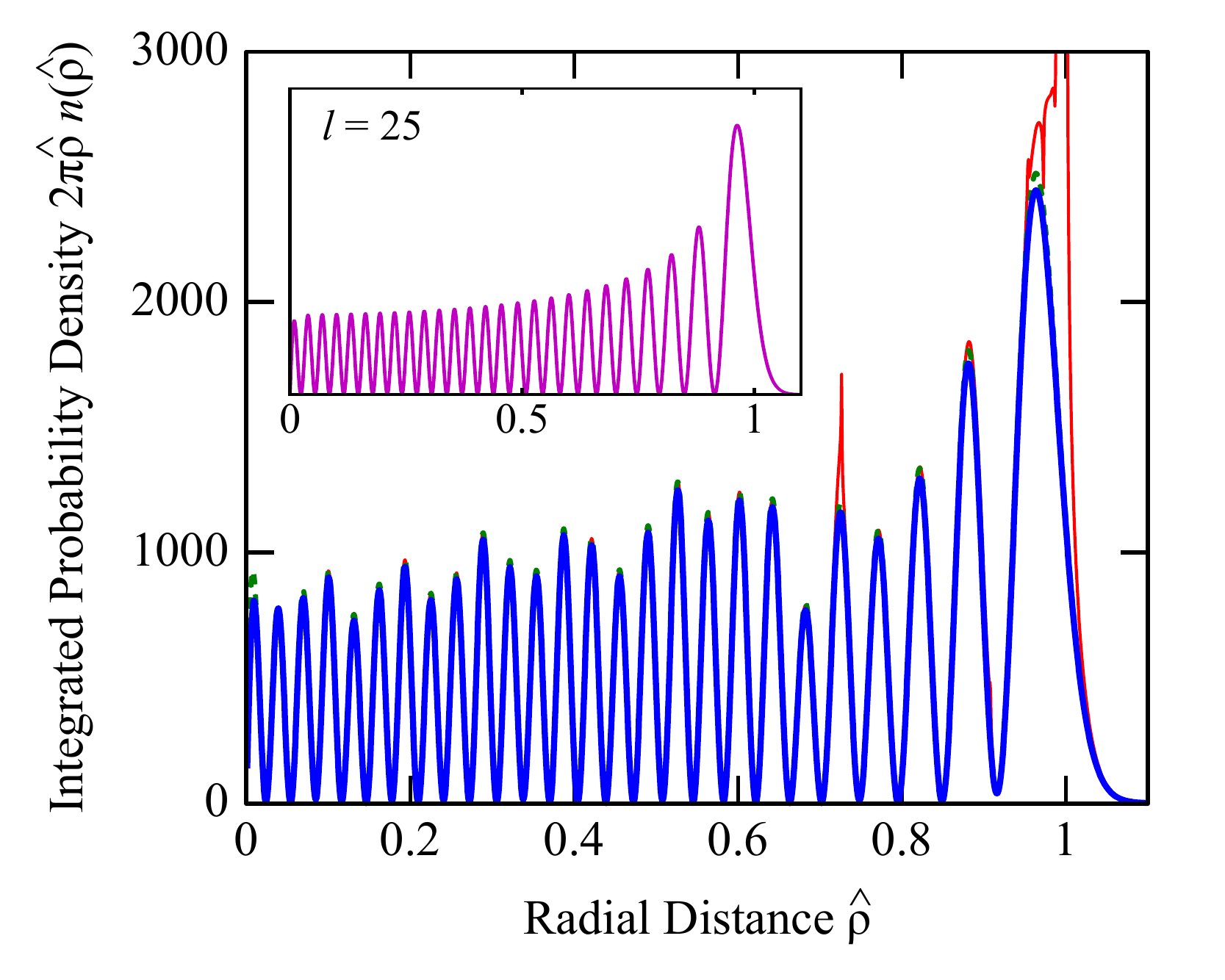}
\caption{\label{eps51:densityprofile}
Radial profile plots of the integrated charge density for energy $\epsilon = 51.01$, at a distance $\hat z = 3.3$ from the source (bottom edge in Figure~\ref{eps51:densitymap}), using the same units as in Figure~\ref{eps50:densityprofile}.  Thin red curve: semiclassical approximation; dashed green curve: uniform approximation; bold blue curve: quantum result.  The density profile $|\psi_{l0}(\hat\rho)|^2$ (\ref{eq:Appendix12}) of the dominant scattering channel (Landau level $l = 25$) is shown in the inset for comparison.}
\end{figure}

From a semiclassical perspective, choosing a near-threshold energy value $\epsilon$ implicates slow convergence of the series (\ref{eq:Semi7}) for the approximate wave function $\psi_\text{sc}(\mathbf r)$ (\ref{eq:Semi1}).  For such values of $\epsilon$, the phase $\Phi_\nu$ (modulo $2\pi$) changes only slowly with each trajectory $\nu$, so the terms in the series (\ref{eq:Semi7}) alternate over long periods, leading to large fluctuations in the partial sums.  Hence, it is necessary to include many classical trajectories in the summation to obtain accurate results in the semiclassical approximation.  Nevertheless, we observe that for a sufficiently large sample size of classical orbits, the trajectory-based methods (in particular, the uniform approximation) are able to reproduce the at times counterintuitive results of the quantum simulation.

For comparison with the results of the previous section, we again provide maps of the particle density $n(\hat\rho,\hat z)$ on a section of the $\hat\rho-\hat z$ plane, calculated using the primitive semiclassical and uniform approximations (left and center panel) as well as the quantum result (\ref{eq:Quantum3}) (right panel in Figure~\ref{eps51:densitymap}).  The caustic surfaces which structure the maps for $\epsilon=50$ (Figure~\ref{eps50:densitymap}) are barely discernible now.  Rather, the density maps now present as a sequence of stripes parallel to the magnetic field direction.  From the quantum perspective, the uniformity of the image is due to the dominance of the near-threshold scattering channel, which imprints its lateral profile onto the density distribution.  It is remarkable that the trajectory-based approximations perform so well in what is essentially a rendering of a single quantum eigenstate in the radial direction.
%
%
\begin{figure}[p]
\includegraphics[width=0.75\textwidth]{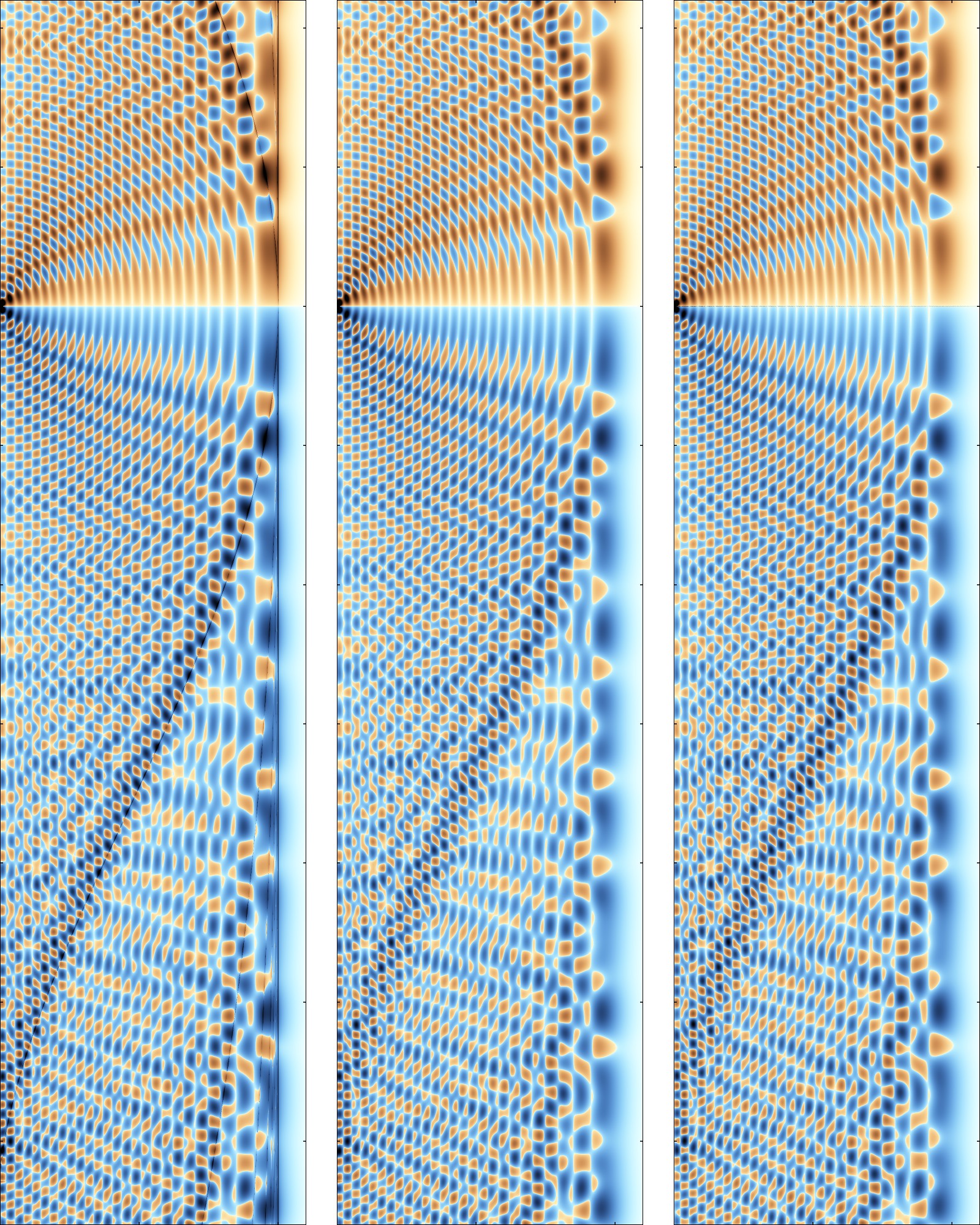}
\caption{\label{eps51:currentmap}
Integrated current density in field direction $2\pi\hat\rho j_z(\hat\rho, \hat z)$ in the $\hat\rho - \hat z$ plane for energy $\epsilon = 51.01$.  Semiclassical calculation (left panel), uniform approximation (center panel), quantum result (right panel).  Parameters as in Figure~\ref{eps51:densitymap}.  The conspicuous ``checkerboard pattern'' of upward (red) and downward (blue) currents indicates widespread prevalence of ``backflow'' toward the source.}
\end{figure}
%
%
\begin{figure}[t]
\includegraphics[width=0.75\textwidth]{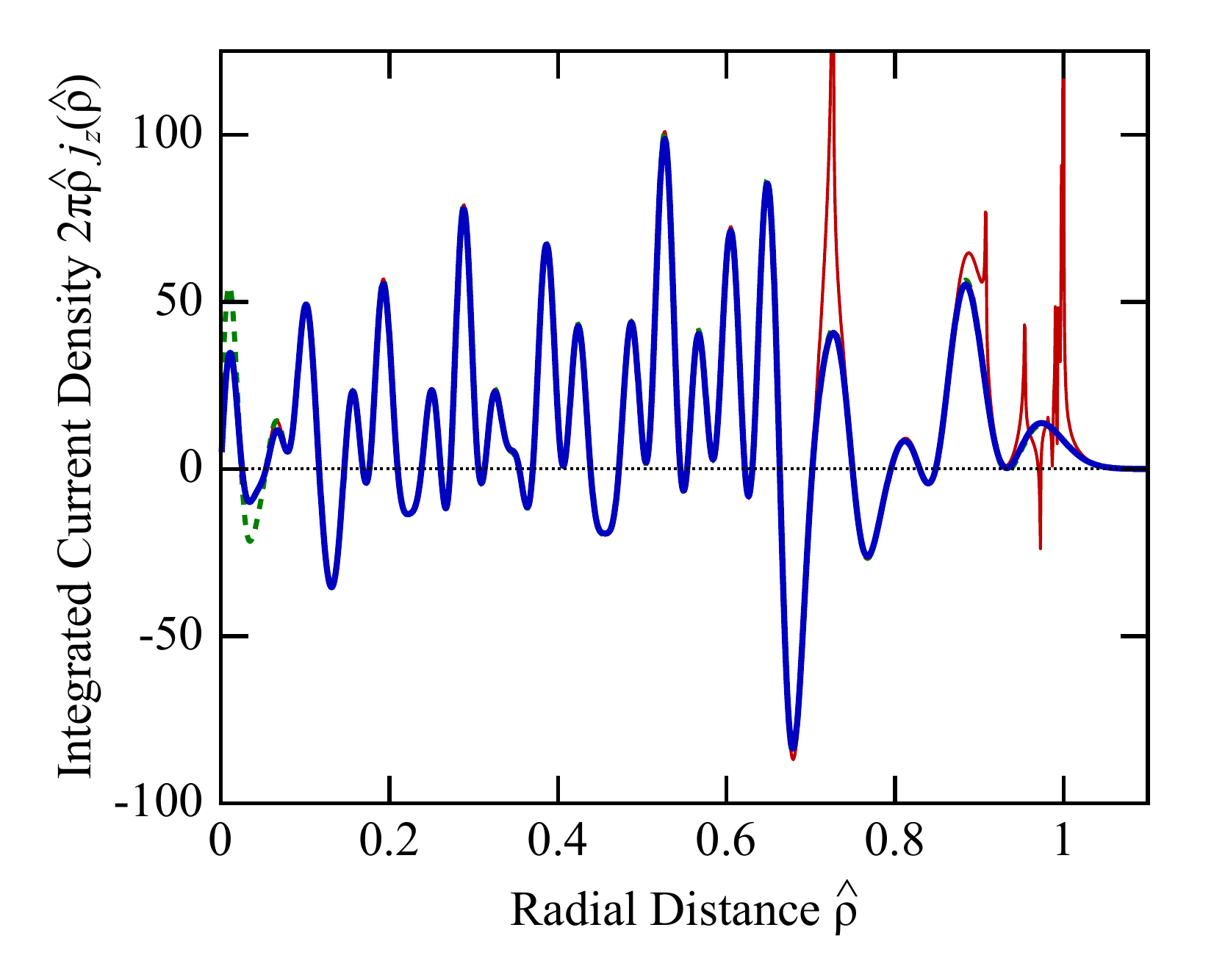}
\caption{\label{eps51:currentprofile}
Radial profiles of the integrated current density $2\pi\hat\rho j_z(\hat\rho)$ for energy $\epsilon = 51.01$; other parameters and units as in Figure~\ref{eps50:currentprofile}.  The semiclassical method (thin red curve), uniform approximation (dashed green curve), and exact quantum result (bold blue curve) all agree in their prediction of sizeable ranges and magnitudes of backflow, indicated by negative current values.}
\end{figure}

To study the density distribution quantitatively, we simulated a profile of the integrated radial density, taken at the bottom edge ($\hat z = 3.3$) of the maps, and plotted it in Figure~\ref{eps51:densityprofile}.  The peak density, near the outer classical limit, is more than thirty times larger than in the ``regular'' case $\epsilon =50$ (Figure~\ref{eps50:densityprofile}), and the density profile strongly resembles the density distribution $|\psi_{l0}(\hat\rho)|^2$ for the Landau level $l =25$ (shown in the inset) as predicted, even though the height of the individual maxima is modulated in the actual profile.  The semiclassical (thin red curve) and uniform approximation (green dashed curve) again reproduce the shape of the quantum result (bold blue curve), save for their known failures at the caustics.

Unlike for $\epsilon = 50$, the current density map $j_z(\hat\rho,\hat z)$ associated with the electron wave near threshold bears little resemblance to the charge density map $n(\hat\rho, \hat z)$. We calculated $j_z(\hat\rho,\hat z)$ in the $\hat\rho - \hat z$ plane using the quantum result (\ref{eq:Quantum5}) and the trajectory-based approximations (\ref{eq:Semi9}) akin to Figure~\ref{eps50:currentmap}.  The result, shown in Figure~\ref{eps51:currentmap}, exposes a distinctive ``checkerboard'' pattern.  This pattern is partly defined by a sequence of node lines running parallel to the field axis which clearly trace back to the zeroes of the wave function $\psi_{l0}(\hat\rho)$ for the dominant scattering channel ($l = 25$), prominent in the density plot (Figure~\ref{eps51:densitymap}).  However, now the caustics are conspicuous in the images, and most strikingly, a rapid succession of upward (red) and downward (blue) current areas covers the entire plane.  While the net flux of particles still leads away from the source, the current distribution is no longer uniform in direction, and almost as likely to point toward the source as away from it.  While it is hard to pinpoint the origin of this ``backflow'' phenomenon, similar effects have been observed before in the dynamics of quantum wave packets \cite{Muga1999a}, albeit not on such a massive scale as here.  Since detectors typically are designed to absorb particles, the presence of backflow has profound implications for the quantum measurement process.
%
%
\begin{figure}[p]
\includegraphics[width=0.75\textwidth]{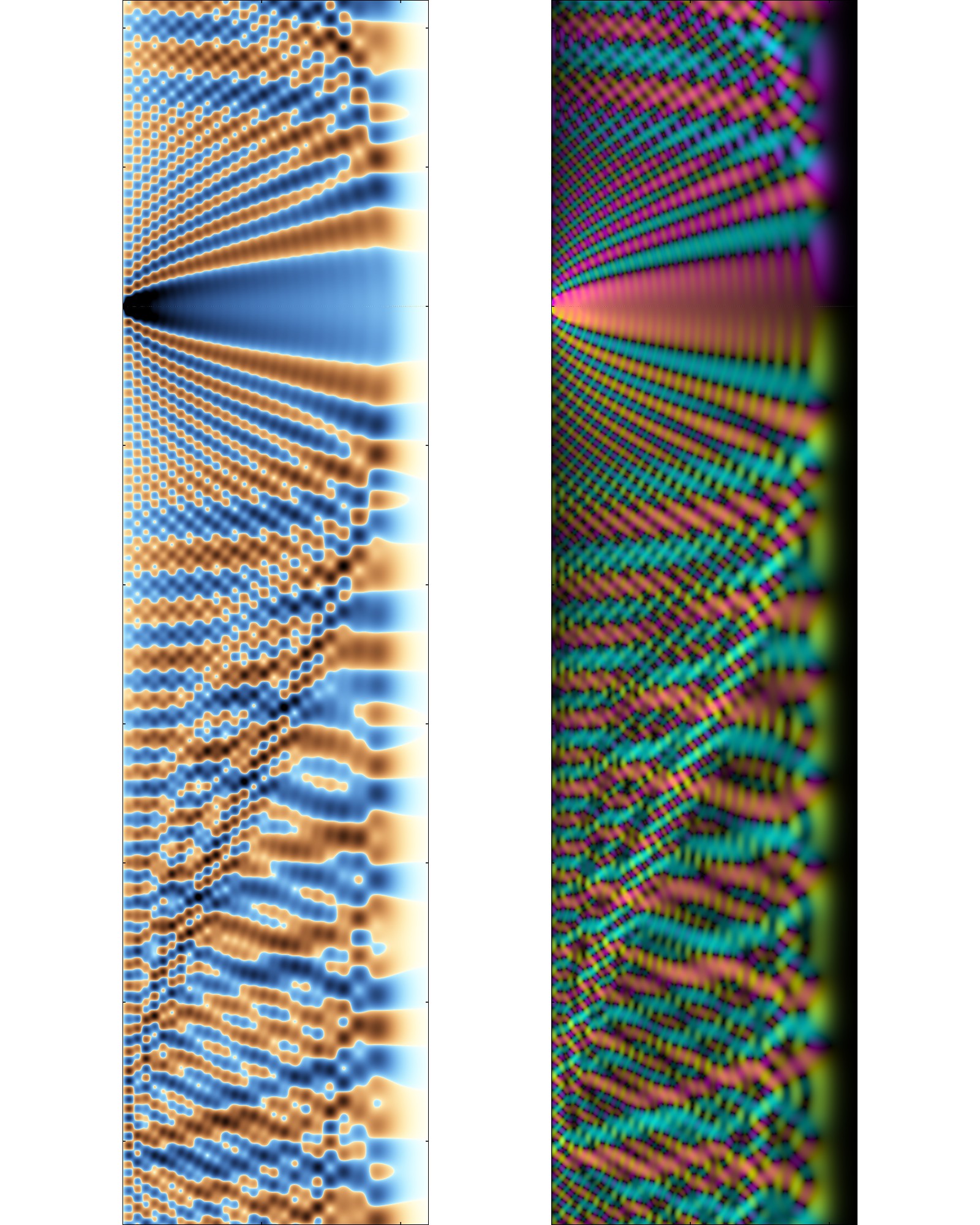}
\caption{\label{eps51:currentfield}
Integrated current density vector $2\pi\hat\rho \mathbf j(\hat\rho, \hat z)$ in the $\hat\rho - \hat z$ plane near threshold ($\epsilon = 51.01$).  Left panel:  The component $j_\rho(\hat\rho, \hat z)$ shows radial flows away (blue) or toward (red) the symmetry axis.  Right panel:  Color-coded current vector field, with brightness indicating magnitude, and hue representing direction (cf.~Figure~\ref{eps50:currentfield}).  Note that the same underlying pattern structures both maps.}
\end{figure}

For a quantitative comparison, we plot the current density profile $j_z(\hat\rho,\hat z)$ in units of $mk^3 / (16\pi^2\epsilon^2\hbar^3)$ at a distance $\hat z = 3.3$, corresponding to the bottom edge of the maps in Figure~\ref{eps51:currentmap}, just as we did before in the regular case $\epsilon = 50$ (Figure~\ref{eps50:currentprofile}).  The result is shown in Figure~\ref{eps51:currentprofile}.  Apart from the known failures near caustics, the orbit-based approximations again accurately trace the quantum current density.  Compared with Figure~\ref{eps50:currentprofile}, the magnitude of the current has increased about fourfold, much less than the corresponding increase in particle density.  The plot demonstrates that backflow (negative values of the current) is a common occurrence, and that the reverse and forward flow are of comparable magnitude.

To complete our comparison with the previous case $\epsilon = 50$, we finally also examine the quantum current in radial direction $j_\rho(\hat\rho, \hat z)$ and the current field $\mathbf j(\hat\rho, \hat z)$ associated with the radial and parallel components of the current for $\epsilon = 51.01$.  The results are displayed in Figure~\ref{eps51:currentfield}, using the same color coding scheme as in Figure~\ref{eps50:currentfield}.  In the radial current map (left panel), we observe a pattern of intense, roughly horizontal bands of inward (red) and outward currents (blue) that almost exactly repeats in the plot of the current field as teal and red patterns (right panel).  This means that the radial component $j_\rho(\hat \rho, \hat z)$ dominates the current distribution:  The particles flow roughly back and forth from the symmetry axis, perpendicular to the magnetic field, with a small and variable contribution in field direction that gives rise to the checkerboard pattern in Figure~\ref{eps51:currentmap}.  In fact, transport parallel to the magnetic field axis (green and purple) is almost exclusively limited to the outermost range of motion, $\hat\rho \approx 1$.

\subsection{Exploring the classical limit}

Another avenue of inquiry is the behavior of the charge in the magnetic field if its energy $E$ is large compared to the typical energy quantum $\hbar\omega_L$.  Its wave function (\ref{eq:Quantum2}) then has contributions from many open scattering channels, and by the correspondence principle, we would expect that classical dynamics emerges from the quantum solution.  However, since this problem has no meaningful classical limit, it is of interest to study the features of the density and current distributions obtained for large values of $\epsilon$.  For our model simulations, we use $\epsilon = 500$, a choice for which the quantum calculation still remains easily managable.

\begin{figure}[t]
\includegraphics[width=\textwidth]{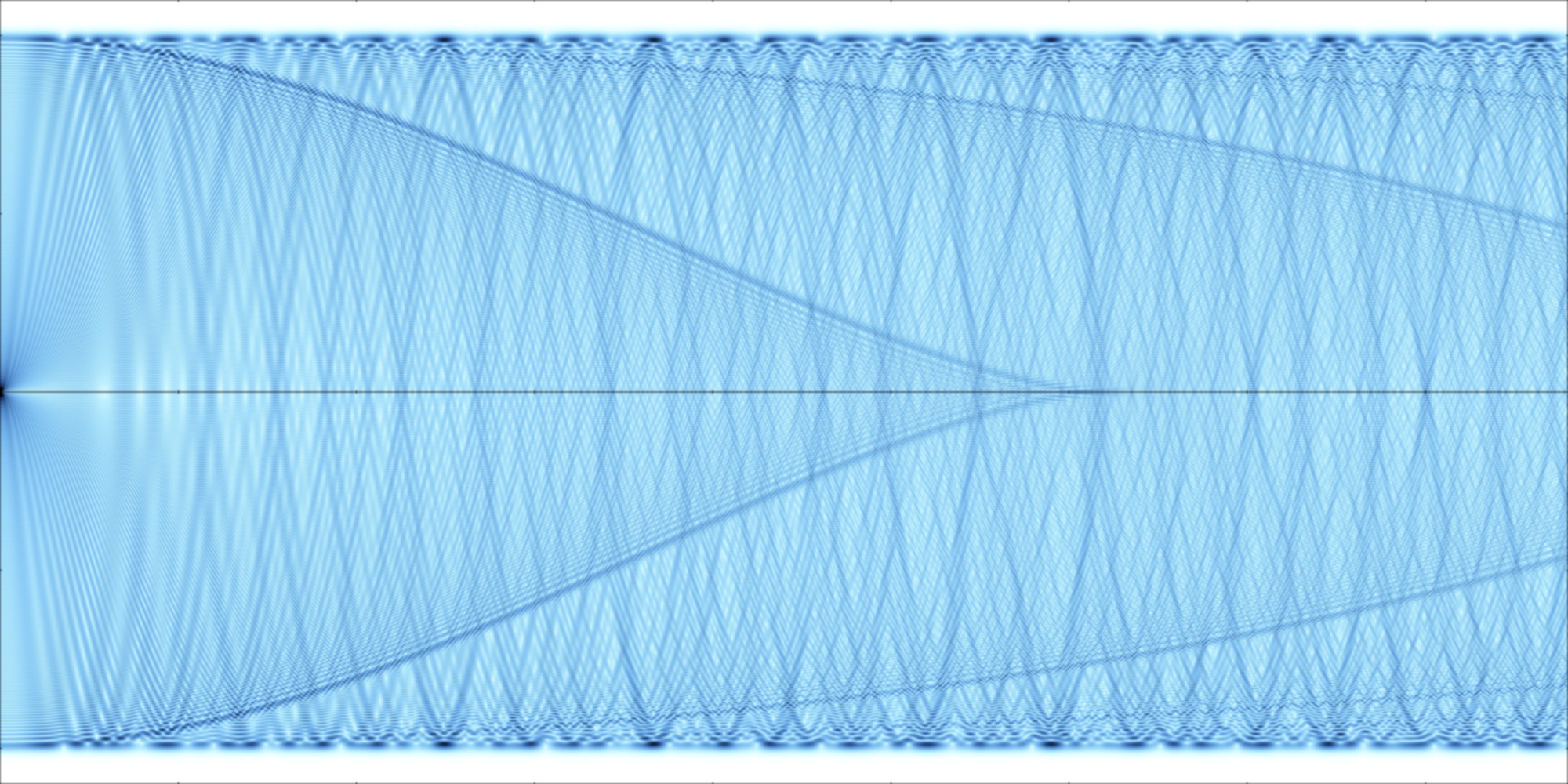}
\caption{\label{eps500:densitymap}
Integrated charge density $2\pi\hat\rho n(\hat\rho, \hat z)$ in a magnetic field (oriented horizontally) along a radial cut in the $\hat\rho - \hat z$ plane, for energy $\epsilon = 500$, evaluated using the quantum result Eq.~(\ref{eq:Quantum2}).  Canvas dimensions are $0 \leq \hat\rho \leq 1.1$ and $0 \leq \hat z \leq 4.4$, with the source located at the center of the left edge.  Dark spots indicate high density.}
\end{figure}

\begin{figure}[t]
\includegraphics[width=\textwidth]{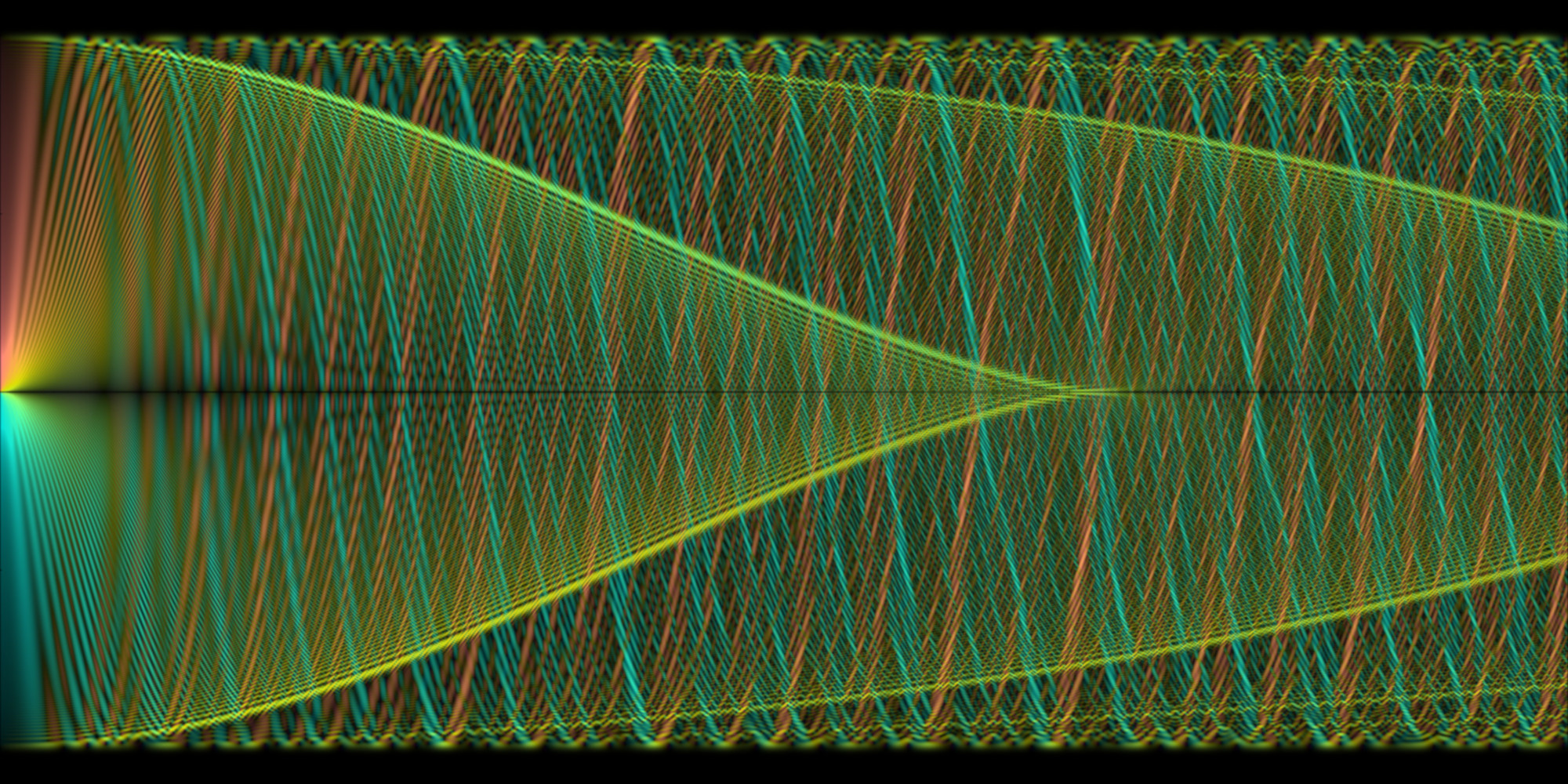}
\caption{\label{eps500:currentfield}
Integrated current density vector field $2\pi\hat\rho \mathbf j(\hat\rho, \hat z)$ corresponding to the density map shown in Figure~\ref{eps500:densitymap}.  Brightness corresponds to the magnitude of $\mathbf j$, while the color hue encodes its direction (red -- upwards, teal -- downwards, yellow -- to the upper left, green - to the lower left).  Current flow occurs at the caustics, as well as along ``arcs'' that transport charge inward and outward.}
\end{figure}

Figure~\ref{eps500:densitymap} shows the integrated density distribution in this case, while Figure~\ref{eps500:currentfield} displays a map of the current density field of the charge.  The color coding in the two images follows the same model as in Figures~\ref{eps50:densitymap} and~\ref{eps50:currentfield}, respectively.  At this high energy, individual interference fringes are too densely spaced to be resolved in these images, save for the outermost range of classical motion \footnote{To avoid Moir\'e effects, much larger maps were created and filtered before downsampling.}.  Virtually all of the patterns seen are emergent ``superstructure.''  Both images clearly display the first few nested caustics as curves of enhanced density and current, accompanied by supernumerary fringes due to the interference  of the incoming and reflected path at the turning surface.  The presence of the caustics is well understood from a semiclassical perspective.  However, the images also show another type of superstructure, a multitude of thin ``arcs'' of various intensity that carry charge from the symmetry axis at the center of the distribution toward the limits of the classical range of motion and back, crossing through the caustics on their way.  These arcs, descendants of the wide bands of current seen in the examples before for lower energies (Figures~\ref{eps50:currentfield} and \ref{eps51:currentfield}), funnel the current away from the source.
%
%
%
%

The arc structure resembles in appearance the ``quantum scars'' frequently observed in billiard problems \cite{Heller1984a,Berry1989a, Wilkinson1996a}, which are linked to classical periodic orbits in such bound systems.  However, the dynamics of a charge in a magnetic field is a scattering problem, with open trajectories, so periodic orbits are absent.  Moreover, the motion is classically integrable, unlike the chaotic dynamics underlying quantum billiards, which renders the significance of the arc pattern even more mysterious.  We have examined maps of varying energy $\epsilon$ like those shown in Figures~\ref{eps500:densitymap} and \ref{eps500:currentfield}, and have found empirically that arcs are present at all energies, and that individual arcs persist as the energy $\epsilon$ is increased from a Landau level threshold to the following threshold, heading slowly away from the source while undergoing fluctuations in intensity.  As the energy sweeps over this interval, it appears that a new arc emerges closest to the source (leftmost arc in the images shown).

\section{Conclusion}
\label{sec:Conclusion}

We conclude with a brief summary of our observations, and an outlook discussing their experimental confirmation.  Notwithstanding the simplicity of the setup, the system harbors interesting behavior and complex features that ultimately trace back to its position at the junction between scattering motion common to open systems, and the periodic motion typical of bound systems.  At the classical level, this complementarity manifests itself in the presence of an infinite number of orbits connecting the source to any destination inside the classically allowed domain $\hat\rho < 1$, and as a consequence, an infinite classical density $n_\text{cl}(\mathbf r)$ that renders the stationary source problem ill-defined.  Thanks to interference between trajectories, the semiclassical model is able to lift this global singularity, and returns finite values for the wave function $\psi_\text{sc}(\mathbf r)$ and derived quantities like the semiclassical density $n_\text{sc}(\mathbf r)$ and current density $\mathbf j_\text{sc}(\mathbf r)$, except for a regular sequence of energy values $E_l = (2l +1) \hbar\omega_L$ ($l =0,1,2,\ldots$) where the divergence persists.  The semiclassical results take the form of conditionally convergent sums over the contributions of the individual orbits.  The same singularities occur in the energy Green function $G(\mathbf r,\mathbf 0;E)$, the quantum mechanical solution of the problem, where they are identified as the Landau levels, the discrete eigenenergies of a charge confined to a plane perpendicular to the magnetic field.  In general, the quantum solution is a rapidly converging sum over scattering channels, combinations of plane waves in field direction with the various radial eigenstates associated with the Landau levels.  Using extensive simulations, we inquired into the properties of the charged particle wave emerging from the source, and compared the different approaches.  We found that the semiclassical results, despite their tenuous convergence properties, were generally in excellent agreement with the quantum solution, in particular when the technique was extended to incorporate uniform approximations based on Airy functions.  The charge and current distributions display rich detail and a hierarchy of features, from individual interference fringes to various ``superstructure.''  A nested set of shells ending in cusps on the symmetry axis is easily identified with the classical caustics of the system, whereas pervasive arc-like features in the density and current maps, reminiscent of the quantum ``scars'' occurring in closed chaotic systems, defy easy explanation.  The simple energy dependence of the total current $J(E)$ emitted by the source sharply contrasts with the complex spatial distribution of the current density $\mathbf j(\mathbf r)$ which exhibits intriguing phenomena like ``backflow'' toward the source.

Experimental verification of the results and simulations presented in this paper is a challenging task, and likely not possible with state-of-the-art equipment.  The established standard for recording the current distribution of charged particle waves propagating in external fields is photodetachment microscopy \cite{Blondel1996a,Blondel1999a,Blondel2010a}.  In electric field experiments, the interference images obtained have diameters $d$ of several mm, and individual fringes are resolved if their spacing $R$ exceeds about 100 $\mu$m \cite{Blondel2001a,Blondel2005a,Pelaez2009a,Pelaez2011a}.  In a {\sl magnetic} field, the size of the electron distribution is given by $d = v_0 / \omega_L$ (\ref{eq:Classical7}), and the composition of the interference pattern itself depends only on the dimensionless energy $\epsilon = E/(\hbar\omega_L)$ (\ref{eq:Quantum1}).  The results of Section~\ref{sec:Results} indicate that the spacing of individual fringes is approximately their ratio $d/\epsilon = \lambda / \pi$, which depends solely on the De Broglie wavelength $\lambda = h /\sqrt{2mE}$ of the charge.  Hence, the energy should not exceed $E = 2\hbar^2/(mR^2) \approx 1.5\cdot 10^{-11}$~eV, an extraordinarily small value compared to the energies used in electric field experiments ($10^{-5}$~eV).  Resolution $R$ and the radius $d$ of the pattern then fix the magnetic field, ${\cal B} = 4\hbar/(qRd)$.  For $d = 5$~mm, the field strength thus obtained is minuscule, ${\cal B} \approx 5\cdot 10^{-9}$~T.  Therefore, observation of individual interference fringes is currently not feasible.

Experimentally accessible values for $E$ and $\cal B$ lead to fringe separations $R$ that are much smaller than the resolution of the instrument.  In this case, the photodetachment image will be governed by the superstructure imposed on the electron distribution.  Because their location and strength shifts significantly between two Landau level thresholds, the variation in $\epsilon = E/(\hbar\omega_L) = d/R$ must be below unity in order to record the arc-like structures.  Meeting this condition requires uncertainties in the energy distribution of the electrons, and drifts of the magnetic field, well below the ratio $1/\epsilon = R/d$.  For energies in the $\mu$eV range, the uncertainties must be of order $10^{-5}$, a very difficult feat to achieve.  In addition, the unavoidable velocity spread of the negative ions in the beam \footnote{The motional electric field related to the average ion velocity $\overline{\mathbf v}_\text{ion}$ can be eliminated by applying a suitable external field $\bm{\mathcal E} = - \overline{\mathbf v}_\text{ion}  \times \bm{\mathcal B}$.} causes varying motional electric fields $\bm{\mathcal E}_\text{mot}$ that must be kept small in comparison to the Lorentz force on the electron, ${\mathcal E}_\text{mot} \ll v_0{\mathcal B}$.  Thus, unless conditions are very precisely controlled, the spread in $\epsilon$ will wash out the modulations in the image due to the arc-like features.  In comparison, the location of the caustic surfaces is rather insensitive to small changes in the parameters, and they should therefore be most easily recognized in photodetachment images as bands of increased intensity, as discussed earlier by Berry \cite{Berry1981a}.

\begin{acknowledgments}
We appreciate helpful discussions with Kevin Mitchell and Tobias Kramer.  A.~G.\ thanks Bard College for their hospitality.  This project has been financially supported through California State University Long Beach and Bard College.
\end{acknowledgments}

\appendix*

\section{Energy Green Function of a Charge in a Magnetic Field}
\label{sec:Appendix}

In this appendix, we briefly discuss the relationship between the energy Green function $G(\mathbf r, \mathbf 0; E)$ and the current emitted by a monochromatic point source located at the origin, establish the particle density and currents associated with a free-particle source as useful quantities for scaling the results in a magnetic field, and outline a method to find the Green function in a homogeneous magnetic field itself.

\subsection{Green Function and Current}
\label{sec:Appendix::Current}

The energy Green function $G(\mathbf r, \mathbf 0; E)$ associated with a stationary Hamiltonian $\mathcal H(\mathbf r, \mathbf p)$ is a solution of the inhomogeneous Schr\"odinger equation with energy $E$:
\begin{equation}
\label{eq:Appendix1}
\left[E - \mathcal H(\mathbf r,\mathbf p)\right] G(\mathbf r, \mathbf 0; E) \,=\, \delta(\mathbf r) \;.
\end{equation}
It can be shown \cite{Bracher1999a} that the Green function is uniquely defined only if $E$ is not part of the energy spectrum of $\mathcal H$.  If $E$ is one of the discrete eigenenergies of the Hamiltonian, $G(\mathbf r, \mathbf 0; E)$ does not exist at all, whereas an entire space of solutions is available if $E$ is a member of the continuous spectrum of $\mathcal H$.  This situation occurs in scattering problems without a confining potential, where the particles are free to leave the system.  For our magnetic field problem, an interesting ``mixed'' case arises where the motion is bound in the $x$--$y$ plane, leading to discrete Landau energy levels associated with the magnetic field, but particles are traveling freely along the $z$--axis, with a continuous range of energies.

The degeneracy in the solutions for Eq.~(\ref{eq:Appendix1}) in the continuous spectrum corresponds physically to the ability to impose boundary conditions on scattering wave functions.  Since we study a source of particles, we are interested in the retarded Green function, the particular outgoing wave solution that carries particles away from the source.  In the vicinity of the origin, this wave invariably gains the simple isotropic characteristics of a spherical $s$--wave \cite{Bracher1999a}, and we therefore identify it with the radially spreading trajectory field employed in the semiclassical study.

The usual expressions for the particle density $n_\text{qm}(\mathbf r)$ (\ref{eq:Quantum3}) and the current density $\mathbf j_\text{qm}(\mathbf r)$ (\ref{eq:Quantum4}) hold in the outgoing electron wave.  To find the total current $J(E)$ emitted by the source, we note that the solutions to Eq.~(\ref{eq:Appendix1}) obey a modified equation of continuity that includes a source term at the origin \cite{Kramer2002a}:
\begin{equation}
\label{eq:Appendix2}
\mathop{\text{div}} \mathbf j_\text{qm}(\mathbf r) \,=\, 
-\frac2\hbar \Im\left[ G(\mathbf r, \mathbf 0; E) \right] \delta(\mathbf r) \;.
\end{equation}
While the Green function $G(\mathbf r, \mathbf 0; E)$ is itself divergent near the source, its imaginary part has a well-defined limit which directly yields the total current $J(E)$.  We apply Gauss' theorem to Eq.~(\ref{eq:Appendix2}), and integrate over a surface enclosing the origin to find:
\begin{equation}
\label{eq:Appendix3}
J(E) \,=\, -\frac2\hbar \left( \lim_{\mathbf r \rightarrow \mathbf 0} \Im\left[ G(\mathbf r, \mathbf 0; E) \right] \right) \;.
\end{equation}

\subsection{Free-Particle Green Function}
\label{sec:Appendix::Free}

In the absence of electric and magnetic fields, i.~e., for the free-particle Hamiltonian $\mathcal H_\text{free} = \mathbf p^2 / (2m)$, the energy Green function becomes an outgoing spherical $s$--wave.  A detailed analysis yields \cite{Bracher1999a}:
\begin{equation}
\label{eq:Appendix4}
G_\text{free}(\mathbf r, \mathbf 0; E) \,=\, - \frac m{2\pi\hbar^2} \frac{e^{ikr}}r \;,
\end{equation}
where $k = mv_0/\hbar$ is the wave number of the electron.  For comparison, we state the particle density $n_\text{free}(r)$ and current density $\mathbf j_\text{free}(r)$ for the free-particle source, using the cyclotron length units introduced in Section~\ref{sec:Classical}.  We set $\hat r = \omega_L r / v_0$ and note that the dimensionless quantity $kr = 2\epsilon{\hat r}$ becomes a function of the dimensionless energy $\epsilon = E/(\hbar\omega_L)$.  Then,
\begin{equation}
\label{eq:Appendix6}
n_\text{free}(\hat r) \,=\, \frac{m^2 k^2}{16\pi^2\epsilon^2\hbar^4} \frac 1{\hat r^2} \;,
\end{equation}
while the magnitude of the current density is,
\begin{equation}
\label{eq:Appendix7}
j_\text{free}(\hat r) \,=\, n_\text{free}(\hat r) v_0 \,=\, \frac{m k^3}{16\pi^2\epsilon^2\hbar^3} \frac 1{\hat r^2} \;.
\end{equation}
Finally, applying Eq.~(\ref{eq:Appendix3}) to the Green function yields the total current $J_\text{free}(E)$:
\begin{equation}
\label{eq:Appendix8}
J_\text{free}(E) \,=\, \frac{mk}{\pi\hbar^3} \;.
\end{equation}
The characteristic growth of the source efficiency with the square root of the energy is known as Wigner's law \cite{Wigner1948a}.

\subsection{Finding the Magnetic Green Function}
\label{sec:Appendix::Magnetic}

Finally, we briefly outline a derivation of the energy Green function in a uniform magnetic field.  A detailed discussion, including an alternative approach to obtain $G(\mathbf r, \mathbf 0; E)$, is found in Ref.~\onlinecite{Kramer2005a}.

We first note that the Hamiltonian $\mathcal H$ is separable into two commuting parts, a two-dimensional operator $\mathcal H_\perp$ that describes the dynamics of the electron in the plane perpendicular to $\bm{\mathcal B}$, and $\mathcal H_\|$, a one-dimensional free-particle Hamiltonian that describes the drift along the magnetic field axis:
\begin{equation}
\label{eq:Appendix9}
\mathcal H \,=\, \frac 1{2m} (\mathbf p - q\mathbf A)^2 \,=\, \mathcal H_\perp + \mathcal H_\| \;,
\end{equation}
where $H_\| = p_z^2 / (2m)$.  $\mathcal H_\perp$ is closely related to the two-dimensional quantum harmonic oscillator:
\begin{equation}
\label{eq:Appendix10}
\mathcal H_\perp \,=\, \frac 1{2m} \left( p_x^2 + p_y^2  \right) - \omega_L L_z + \frac{m\omega_L^2}2 \left( x^2 + y^2 \right) \;.
\end{equation}
($L_z$ denotes the angular momentum component in $z$--direction.)  Since the electron motion is bound in lateral direction, the spectrum of $\mathcal H_\perp$ is discrete, and given by the Landau levels:
\begin{equation}
\label{eq:Appendix11}
E_{l\mu} \,=\, (2l+1) \hbar\omega_L \;, \qquad (l = 0,1,2,\ldots) \;,
\end{equation}
i.~e., $\epsilon_{l\mu} = 2l+1$.  Each Landau level has infinite degeneracy, as the magnetic quantum number of the electron $\mu = -l, -l+1,-l+2, \ldots$ can take any integer value greater or equal to $-l$.  However, only the eigenstates $\psi_{l\mu}$ with $\mu = 0$ are of interest here, since all others vanish at the source location.  In polar coordinates, the corresponding normalized eigenfunctions $\psi_{l0}(\hat\rho)$ read, using again the dimensionless length $\hat\rho = \omega_L \rho / v_0$:
\begin{equation}
\label{eq:Appendix12}
\psi_{l0}(\hat\rho) \,=\, \sqrt{\frac{m\omega_L}{\pi\hbar}} L_l(2\epsilon \hat\rho^2) e^{-\epsilon \hat\rho^2} \;,
\end{equation}
where $L_l(u)$ denotes a Laguerre polynomial \cite{Moore1977a}.  Since $L_l(0) = 1$, $\psi_{l\mu}(0) = [m\omega_L/(\pi\hbar)]^{1/2}\,\delta_{\mu 0}$ holds at the origin $\hat\rho =0$.  Therefore, the completeness relation for the eigenstates of $\mathcal H_\perp$ can be stated as:
\begin{equation}
\label{eq:Appendix13}
\sum_{l, \mu} \psi_{l\mu}(x,y)\psi_{l\mu}(0,0)^* \,=\, \frac{m\omega_L}{\pi\hbar} e^{-\epsilon \hat\rho^2} \sum_{l=0}^\infty L_l(2\epsilon \hat\rho^2) \,=\, \delta(x)\delta(y) \;.
\end{equation}

Physically, the separability of the problem implies that the wave function of the electron is a superposition of scattering waves that occupy different Landau levels $\psi_{l0}$, with the ``remaining'' energy $E_\| = E - E_{l0}$ associated with the drift motion in $z$--direction.  Indeed, if we introduce the one-dimensional free-particle Green function $G_\text{1D}(z,0;E)$ as the solution of $\left[ E - \mathcal H_\| \right] G_\text{1D}(z, 0; E) = \delta(z)$, we find,
\begin{equation}
\label{eq:Appendix15}
\left[ E - \mathcal H \right] \psi_{l0}(\hat\rho) G_\text{1D}(z, 0; E - E_{l0}) \,=\,
\psi_{l0}(\hat\rho) \left[ E - E_{l0} - \mathcal H_\| \right] G_\text{1D}(z, 0; E - E_{l0}) \,=\,
\psi_{l0}(\hat\rho) \delta(z) \;.
\end{equation}
In view of Eq.~(\ref{eq:Appendix13}), the full Green function $G(\mathbf r, \mathbf 0; E)$ has the series representation:
\begin{equation}
\label{eq:Appendix16}
G(\mathbf r, \mathbf 0; E) \,=\, \sum_{l=0}^\infty \psi_{l0}(\hat\rho)\psi_{l0}(0)^* G_\text{1D}(z, 0; E - E_{l0}) \;.
\end{equation}
The remaining task is to find an expression for $G_\text{1D}(z, 0; E_\|)$.  As a free-particle Green function, it must have the form of an outgoing wave in either direction of the $z$--axis (or in the case $E_\|<0$, an evanescent wave).  A simple analysis \cite{Kramer2005a} shows that:
\begin{equation}
\label{eq:Appendix17}
G_\text{1D}(z, 0; E_\|) \,=\, 
\begin{cases}
\frac{m}{i\hbar^2 k_\|} e^{ik_\||z|} & \qquad (E_\|>0) \;, \\
- \frac{m}{\hbar^2 k_\|} e^{-k_\||z|} & \qquad (E_\|<0) \;,
\end{cases}
\end{equation}
where $k_\| = \sqrt{2 m|E_\||} /\hbar$.  We introduce again dimensionless coordinates, combine Eqs.~(\ref{eq:Appendix13})--(\ref{eq:Appendix17}), and finally obtain the full Green function:
\begin{equation}
\label{eq:Appendix19}
G(\mathbf r, \mathbf 0;E) \,=\, \frac{mk}{2\pi\hbar^2} e^{-\epsilon \hat\rho^2} \left( 
\sum_{2l+1 < \epsilon} L_l(2\epsilon \hat\rho^2) \frac{e^{2i\sqrt{\epsilon(\epsilon-2l-1)} |\hat z|}}{i\sqrt{\epsilon(\epsilon - 2l - 1)}} -
\sum_{2l+1 > \epsilon} L_l(2\epsilon \hat\rho^2) \frac{e^{-2\sqrt{\epsilon(2l+1 -\epsilon)} |\hat z|}}{\sqrt{\epsilon(2l + 1 - \epsilon)}} \right) \;.
\end{equation}
The summation runs over all $l=0,1,2,\ldots$.  If the energy coincides with one of the Landau levels $\epsilon_l = 2l+1$, the Green function diverges.

\bibliography{Magnetic2011}

\end{document}